\documentclass[a4paper,fleqn,usenatbib]{mnras}

\usepackage[T1]{fontenc}
\usepackage{ae,aecompl}

\usepackage{graphicx}
\usepackage{amsmath}
\usepackage{amssymb}

\title[Pulsating variables in DDO210]{The pulsating variable star population in DDO210}

\author[Ordo\~{n}ez \& Sarajedini]{
Antonio J. Ordo\~{n}ez$^{1}$\thanks{E-mail: a.ordonez@ufl.edu}
and Ata Sarajedini$^{1}$
\\
$^{1}$Department of Astronomy, University of Florida, 211
  Bryant Space Science Center, Gainesville, FL 32611, USA\\
}

\date{Accepted 2015 October 23. Received 2015 October 20; in original
  form 2015 August 28}

\pubyear{2015}

\begin{document}
\label{firstpage}
\pagerange{\pageref{firstpage}--\pageref{lastpage}}
\maketitle

\begin{abstract}
We have probed the pulsating variable star content of the isolated Local Group
dwarf galaxy, DDO210 (Aquarius), using archival Advanced Camera for
Surveys/$Hubble$ $Space$ $Telescope$ imaging in the
F475W and F814W passbands. We find a total of 32 RR Lyrae stars (24
ab-type; 8 c-type) and 75 Cepheid variables. The mean periods of the ab-type
and c-type RR Lyrae
stars are calculated to be $\langle$P$_{\mathrm{ab}}\rangle = 0.609\pm0.011$ and $\langle$P$_{\mathrm{c}}\rangle = 0.359\pm0.025$ days, respectively. The light curve properties of the
fundamental mode RR Lyrae stars yield a mean metallicity of $\langle$[Fe/H]$\rangle$ =
-1.63$\pm$0.11 dex for this ancient population, consistent with a
recent synthetic colour-magnitude diagram analysis. We find this galaxy to
be Oosterhoff-intermediate and lacking in high-amplitude, short-period
ab-type RR Lyrae, consistent with behavior recently observed for many
dwarf spheroidals and ultra-faint dwarfs in the Local Group. We find a distance modulus of $\umu =
25.07\pm 0.12$ as determined by the RR Lyrae stars, slightly larger
but agreeing with recent distance estimates from the red giant branch tip. We also
find a sizable population of Cepheid variables in this
galaxy. We
provide evidence in favor of most if not all of these stars being
short-period classical Cepheids. Assuming all of these stars to be
classical Cepheids, we find that most of these Cepheids are $\sim$300
Myr old, with the youngest Cepheids being offset from the older
Cepheids and the centre of
the galaxy. We conclude that this may have resulted from a migration of star
formation in DDO210.
\end{abstract}


\section{Introduction}
\label{sec:intro}
Dwarf galaxies are known to be the most numerous type of galaxy in
general. Additionally, dwarf
galaxy accretion has been proposed as one of the major mechanisms in
the formation of massive galaxies, such as the Milky Way (MW)
\citep{mat98}. While the notion of MW halo build-up from systems
resembling modern day dwarf spheroidals (dSph) has encountered
significant contention recently from element abundance ratio comparisons with
the Galaxy \citep{ven04, pri05b} and the Oosterhoff dichotomy
\citep{cat09}, accretion of dwarfs has certainly occurred, with the
Sagittarius dSph providing a smoking gun for such interactions
\citep{iba94}. Studying the stellar populations of Local Group (LG) dwarfs,
especially the old ones, allows us to better constrain the extent to which
these dwarfs have contributed to the Galactic halo. Dwarf
galaxies also present unique astrophysical laboratories to study how
these relatively simple galaxies have evolved in different
environments through their chemical enrichment and star formation 
histories (SFHs). 

Variable stars are important tracers of the histories of dwarf
galaxies. They provide unique insights into
their parental stellar populations through the study of their light
curves. Cepheid and RR Lyrae variables are especially useful in
this context. For instance, classical Cepheids (CCs) are massive,
blue-loop stars that are relatively young, having formed within the
past $\sim$1 Gyr \citep{bon05}. Therefore these stars trace young star
formation events. On the other hand, RR Lyrae stars are ancient
horizontal branch (HB) stars originating from low-mass stars, and thus
trace star formation at ages $\ge$10 Gyr (see Figs. 2 and 3
of \citet{ldz94} for theoretical modeling of RR Lyrae stars on the HB
and \citet{gla08} for an age determination of the youngest known
system harboring RR Lyrae stars). Anomalous
Cepheids (ACs) are thought to represent more intermediate-mass
stars at low metallicities. They may trace intermediate age (1-6 Gyr)
populations or old binary systems \citep{fm12}. The mere presence of any
combination of these stars thus provides constraints on the SFH of the
host system.

In addition to informing the SFH of a stellar population, pulsating
variables also provide insight into the chemical enrichment of such a
population. Most notably, there are strong correlations between the light curve
shapes of fundamental mode RR Lyrae, referred to in the literature as
RRab stars, and their iron abundances, [Fe/H] \citep{jk96, alc00,
  nem13}. Similar relationships for Cepheids have recently been
explored \citep{sza12, kla13}. Therefore, these stars provide a means
to elucidate portions of the chemical enrichment history of a stellar population
without the need for spectroscopic observations. 

The LG dSph/Irr transition-type (dTrans; see \citet{mat98}) galaxy, DDO210 (Aquarius), is one of the
most distant dwarf galaxies in the LG. While distance estimates to
this galaxy have differed significantly over the past few decades,
recent values have converged on a distance of $d\sim 1$ Mpc using the
tip of the red giant branch (TRGB) \citep{col14}. This galaxy is one
of the most isolated in the LG and thus provides an excellent
opportunity to study how such low-mass, isolated dwarfs evolve. 

Recently, \citet{col14} presented a synthetic colour-magnitude diagram
(CMD) analysis of the SFH
of DDO210 using their $Hubble$ $Space$ $Telescope$ ($HST$)/Advanced Camera for
Surveys (ACS) observations (GO-12925; PI:
A. Cole). They find a complex SFH over the lifetime of DDO210
characterized by a long delay before the onset of a major star burst
$\sim$7 Gyr ago and followed by a relative lull in star formation for the past
$\sim$5 Gyr. The age-metallicity relation (AMR) produced by their
analysis reveals very little chemical enrichment throughout its lifetime. 

The goal of this work is to identify and characterize the pulsating
variable stars within DDO210, thereby constraining its evolutionary
history. In Section \ref{sec:obs}, we describe the data
used for this study, and the reduction process performed on them. We
detail the variable star identification and characterization processes
in Section \ref{sec:var}. We present the properties of the RR Lyrae
and Cepheid stars in Sections \ref{sec:rrl} and \ref{sec:ceph}
respectively. A discussion of the properties of these stars within the
context of the SFH of DDO210 is given in Section \ref{sec:disc}, and
the conclusions are presented in Section \ref{sec:conc}.

\section{Observations and Data Reduction}
\label{sec:obs}
The data set used by \citet{col14} is very deep
and covers a time baseline conducive to identifying short-period
variable stars. These observations of DDO210 were originally intended for use in a
detailed SFH analysis for this dwarf, and thus cover a significant portion of the galaxy while reaching
photometric depths to the main-sequence turnoff. The observations
consisting of 22,920 seconds in F475W and 33,480 seconds in F814W were
taken with a cadence well-suited for identifying short-period variable
stars. We retrieved these images from the Mikulski
Archive for Space Telescopes
(MAST)\footnote{\url{http://archive.stsci.edu/}} for use in our study. 

We downloaded the charge-transfer efficiency (CTE) corrected ($*FLC$)
images from MAST, which were also processed through the standard $HST$
pipeline. Bad pixels were then masked and the geometric correction
pixel area maps were applied to each image. Photometry was then
performed using the DAOPHOT/ALLSTAR/ALLFRAME software packages \citep{ste87,
  ste94}. Empirical point-spread functions (PSFs) were constructed using the
brightest isolated stars in one cosmic-ray rejected
reference image for each filter using the corresponding task in DAOPHOT. Aperture corrections were
calculated from bright, isolated stars on each chip. The median aperture
correction for these stars was then applied to all stars. We noticed a
small photometric offset between the WFC1 and WFC2 chips after the
aperture correction, and placed the WFC2 photometry on the same scale
as the WFC1 photometry for consistency. The offsets amounted to 0.1
mag in F475W and -0.05 mag in F814W. We chose to correct WFC2 to the
photometric system on WFC1 after comparison of our CMDs with that of
\citet{col14}, who used the same data set, revealed that the WFC1
photometry agreed with theirs. We attribute this error to the lack of
bright, isolated stars in the WFC2 field with which to obtain an accurate aperture
correction value. Finally,
conversion of the native $HST$ VEGAMAG photometry to the ground-based
Johnson-Cousins B and I filters was accomplished using the
prescription from \citet{sir05}.

\section{Variable star characterization and simulations}
\label{sec:var}
\subsection{Characterization}
\label{sub:char}
In order to identify potential variable stars within the photometry,
we first applied cuts in magnitude and colour in order to narrow our
region of interest. Since the primary goal of this work is to
characterize the pulsating variable star content within the
instability strip, we determined stars within $21$ mag
$<$ m$_{\mathrm{F814W}} < 27$ mag and 0 $<$ m$_{\mathrm{F475W}}
-$m$_{\mathrm{F814W}} <1.5$ mag to be an appropriate region of
interest. Potential variables were identified using a reduced $\chi^2$
defined as:
\begin{equation}\label{eq:varchi}
\chi^2 = \frac{1}{\mathrm N_1 + \mathrm
  N_2}\times \left[\sum_{i=1}^{\mathrm N_1}
  \frac{(m_{1, i} -\bar{m_1})^2}{\sigma_i^2} +
  \sum_{i=1}^{\mathrm N_2} \frac{(m_{2, i} - \bar{m_2})^2}{\sigma_i^2} \right]
\end{equation}
In this case, $m_1$ and $m_2$ are the F475W and F814W magnitudes in
the VEGAMAG photometric system. In an effort to filter spurious
variables, we rejected 3-$\sigma$ outliers from the $\chi^2$
calculation for each light curve. Stars with $\chi^2 \ge 2$ were
flagged as variable candidates. 

\begin{figure}
\includegraphics[width=\columnwidth]{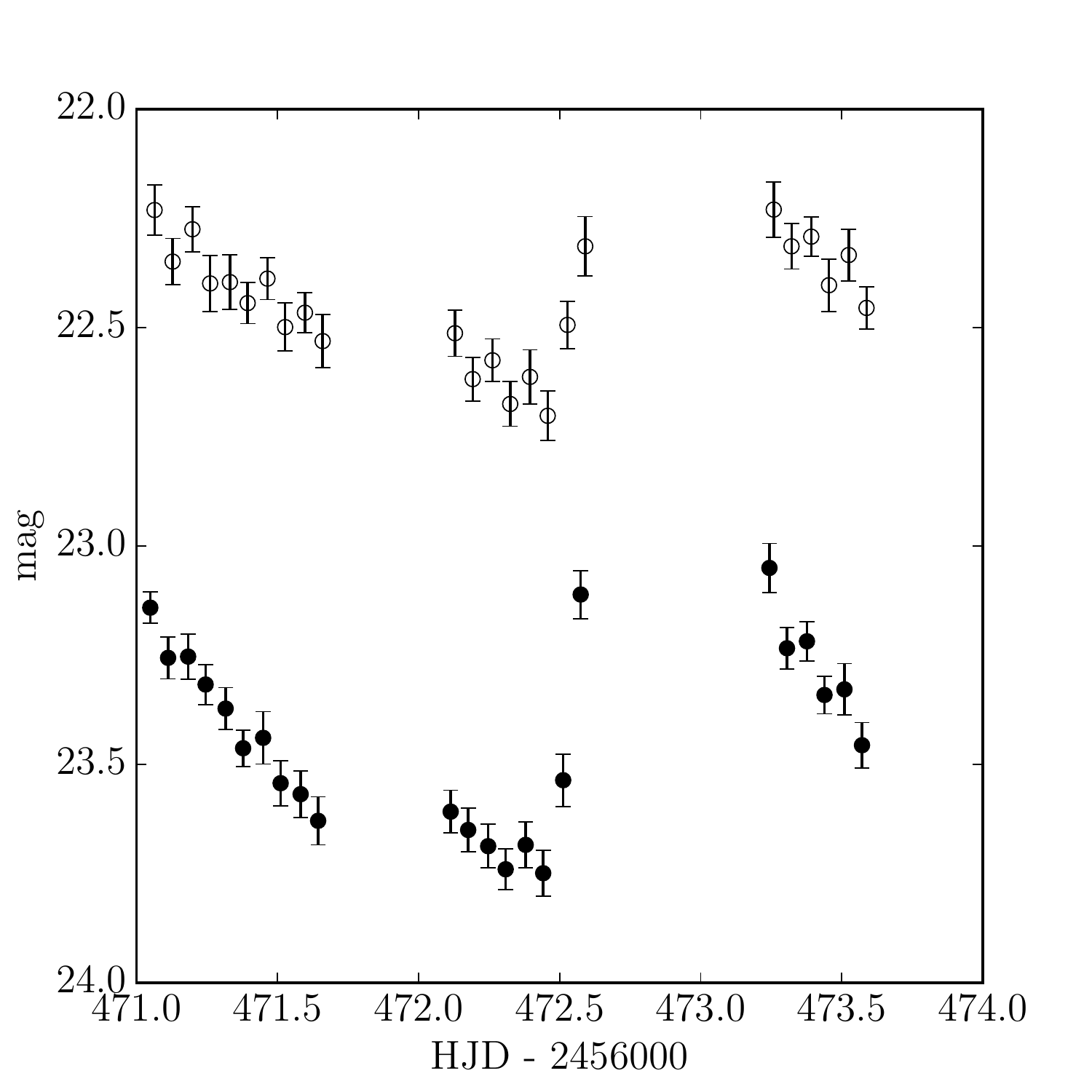}
\caption{\small{The raw, unphased light curve of one variable star
    candidate in our sample. The filled circles represent the F475W
    light curve, while the open circles are the F814W light curve
    data points. \label{fig:raw_obs}}}
\end{figure}
The raw light curve of one of the variable candidates is shown in
Fig. \ref{fig:raw_obs}. This light curve illustrates the high
quality of these data for identifying variable stars. Table
\ref{tbl:timephot} provides the full, time-series photometry for this
star, and the photometry for all variable candidates is available in the
online version of the journal. 
\begin{table}
\caption{An example time-series photometry set for one variable
  star candidate in DDO210. The full set of these
  for all variable star candidates is included in the online
  version of the journal.\label{tbl:timephot}}
\begin{tabular}{ccccc}
\hline
Star ID & Filter & HJD - 2456000 (days) & Magnitude & Magnitude error\\
\hline
V0003 & F475W & 471.0488 & 23.141 & 0.036\\ 
V0003 & F814W & 471.06405 & 22.231 & 0.058\\ 
V0003 & F475W & 471.11204 & 23.256 & 0.048\\ 
V0003 & F814W & 471.12809 & 22.349 & 0.053\\ 
V0003 & F475W & 471.18299 & 23.253 & 0.052\\ 
V0003 & F814W & 471.19824 & 22.275 & 0.052\\ 
V0003 & F475W & 471.24499 & 23.317 & 0.046\\ 
V0003 & F814W & 471.26104 & 22.399 & 0.064\\ 
V0003 & F475W & 471.31594 & 23.372 & 0.048\\ 
V0003 & F814W & 471.33119 & 22.396 & 0.063\\ 
V0003 & F475W & 471.37795 & 23.463 & 0.042\\ 
V0003 & F814W & 471.39401 & 22.444 & 0.047\\ 
V0003 & F475W & 471.4489 & 23.439 & 0.06\\ 
V0003 & F814W & 471.46414 & 22.388 & 0.048\\ 
V0003 & F475W & 471.51092 & 23.543 & 0.052\\ 
V0003 & F814W & 471.52697 & 22.499 & 0.055\\ 
V0003 & F475W & 471.58185 & 23.568 & 0.053\\ 
V0003 & F814W & 471.59709 & 22.466 & 0.046\\ 
V0003 & F475W & 471.64388 & 23.629 & 0.055\\ 
V0003 & F814W & 471.65993 & 22.531 & 0.061\\ 
V0003 & F475W & 472.11366 & 23.608 & 0.049\\ 
V0003 & F814W & 472.1289 & 22.513 & 0.053\\ 
V0003 & F475W & 472.17573 & 23.65 & 0.05\\ 
V0003 & F814W & 472.19179 & 22.618 & 0.05\\ 
V0003 & F475W & 472.24661 & 23.687 & 0.05\\ 
V0003 & F814W & 472.26185 & 22.575 & 0.049\\ 
V0003 & F475W & 472.3087 & 23.74 & 0.047\\ 
V0003 & F814W & 472.32475 & 22.675 & 0.051\\ 
V0003 & F475W & 472.37956 & 23.684 & 0.053\\ 
V0003 & F814W & 472.39481 & 22.613 & 0.062\\ 
V0003 & F475W & 472.44166 & 23.749 & 0.053\\ 
V0003 & F814W & 472.45772 & 22.702 & 0.057\\ 
V0003 & F475W & 472.51251 & 23.536 & 0.06\\ 
V0003 & F814W & 472.52776 & 22.494 & 0.054\\ 
V0003 & F475W & 472.57461 & 23.111 & 0.055\\ 
V0003 & F814W & 472.59067 & 22.314 & 0.068\\ 
V0003 & F475W & 473.24368 & 23.05 & 0.056\\ 
V0003 & F814W & 473.25892 & 22.23 & 0.063\\ 
V0003 & F475W & 473.30586 & 23.234 & 0.048\\ 
V0003 & F814W & 473.32192 & 22.314 & 0.052\\ 
V0003 & F475W & 473.37662 & 23.218 & 0.045\\ 
V0003 & F814W & 473.39187 & 22.292 & 0.045\\ 
V0003 & F475W & 473.43881 & 23.341 & 0.043\\ 
V0003 & F814W & 473.45487 & 22.403 & 0.06\\ 
V0003 & F475W & 473.50957 & 23.328 & 0.059\\ 
V0003 & F814W & 473.52482 & 22.334 & 0.059\\ 
V0003 & F475W & 473.57178 & 23.456 & 0.052\\ 
V0003 & F814W & 473.58783 & 22.455 & 0.048\\ 
\hline
\end{tabular}
\end{table}

Once we identified our set of variable candidates, we then removed
data points from each light curve with anomalously high errors
($\ga 0.1$ mag) to ensure a final sample of high fidelity light
curves. Analysis of these light curves was then performed using
template light curve fitting. This method, based largely off the
technique originated by \citet{lay99}, caters well to data sets with
sparse or
irregular time sampling and large gaps. We use \texttt{RRFIT} \citep{yng12} to
perform the light curve fitting. To summarize the method, \texttt{RRFIT}
performs a robust search through the user defined parameter space of
expected pulsational properties to find the best-fitting light curve
template for each star. The program does so by thoroughly searching
the defined period space and generating potential light curves based
upon the supplied templates at each period
using the genetic algorithm, PIKAIA \citep{cha95}, to optimize each template
to fit the data. Each light curve fit is then ranked according to its reduced
$\chi^2$, and the parameters that minimize this value
are taken to be the best-fitting light curve. 

Given
previous studies of the stellar populations and SFH in DDO210, we
expect there to be a sizable population of RR Lyrae variables as well
as Cepheid variables.  Considering this, we performed a first-pass
with \texttt{RRFIT} on our entire set of variable
candidates searching for periods in the range 0.2--1.0 days, which
allowed us to distinguish between potential faint, short-period stars
and the brighter stars with presumably longer periods. Following this
first pass, we divided the sample into two groups by making a cut in
F814W at 24.25 mag. Variables fainter than this were taken as RR Lyrae
candidates, while brighter variables were assigned as Cepheid
candidates. 

We then ran a second round of \texttt{RRFIT} on each set of variable
candidates, this time tailoring the pulsational paramater space to
match each type of variable. In the case of the Cepheid
candidates, we searched for periods in the range of 0.4--4.0 days,
covering the period distributions for the different subclasses of
Cepheids found in LG dwarf galaxies. We chose to use the RR Lyrae
templates for the Cepheids since the light curve shapes are
similar. For the RR Lyrae candidates, we
searched for periods in the usual range of 0.2--1.0 days.

 We took advantage of the relationship between pulsation amplitudes in
different bandpasses for the RR Lyrae stars as discussed in
\citet{df99} in order to improve the template fitting procedure for
these stars. \citet{df99} provide these relationships in the
Johnson--Cousins system for the B, V, and I filters. In order to derive
similar transformations for the VEGAMAG photometric system, we
used the following technique. First, we generated 4,000
artificial RR Lyrae light curves using the templates in \texttt{RRFIT}. These
artificial light curves were generated with the following constraint
on their amplitudes from \citet{df99}:
\begin{equation} \label{eq:ampconstjc}
A_B = 0.102 + 1.894A_I
\end{equation}
We then transformed these light curves to the VEGAMAG magnitudes in
F475W and F814W using the prescription from \citet{sir05} at each phase in the
light curve. Finally, the amplitudes in F475W and F814W were
calculated for each synthetic light curve, and the following linear relation was
fit to these amplitudes:
\begin{equation} \label{eq:ampconstvgm}
A_{\mathrm{F475W}} = 0.089 + 1.734A_{\mathrm{F814W}}
\end{equation}
Equation (\ref{eq:ampconstvgm}) thus provides a constraint on the
amplitudes of the RR Lyrae candidates in the VEGAMAG photometric
system. We modified \texttt{RRFIT} to implement this constraint in order to
improve the accuracy of the RR Lyrae fitting routine. On the other hand, for
Cepheids it has been known for some time that the amplitude ratio
$A_I / A_V = 0.6$ mag is independent of period or amplitude
\citep{tan97}. However, the findings of \citet{cc89} suggest a
possible dependence of $A_B / A_V$ on period, so we did not implement
any such amplitude constraint on the Cepheid candidates.

Finally, we performed multiple checks on these best-fitting light curve to
assess their validity. For the RR Lyrae stars, the locations in the Bailey
(period-amplitude) diagram was checked to verify that each light curve's
period, amplitude, and mode follow the well-known behavior of RR Lyrae
stars in this space. We manually checked for potential aliasing by
visually examining all light curves produced by
\texttt{RRFIT}. Light curves with periods suspected to be aliased were
checked with the interactive light curve fitting program,
\texttt{FITLC} \citep{man08}. This software operates on the same
principles as \texttt{RRFIT} but provides a GUI for the user to
examine the light curves and the period-$\chi^2$ space. This program
illustrates the best-fitting light curve for each template overplotted
on the data, along with a plot of reduced $\chi^2$ versus period for
each template. In this way, \texttt{FITLC} allows one to explore
different period/template combinations. Occasionally, the automated
fitting routine misclassifies a star or falls victim to a period
alias. Aliased
periods can be compared with other periods lying at a similar $\chi^2$
minimum. If a fit appears anomalous on the Bailey diagram
(e.g. significantly shorter/longer period at a given amplitude
compared with the other RR Lyrae stars on the diagram), or a period-folded light
curve appears to have a large gap, the fit period is usually aliased
with two or more comparable $\chi^2$ minima. Using FITLC, we explored the
different periods with similarly low $\chi^2$ for these problematic
light curves. If one of these other periods yields light curve
parameters that better match the data and/or previously observed
behavior for that type of variable (e.g. period-amplitude relation),
then that period is taken as the correct one. We will hereafter refer
to this procedure as manual fitting.

\begin{figure}
\includegraphics[width=\columnwidth]{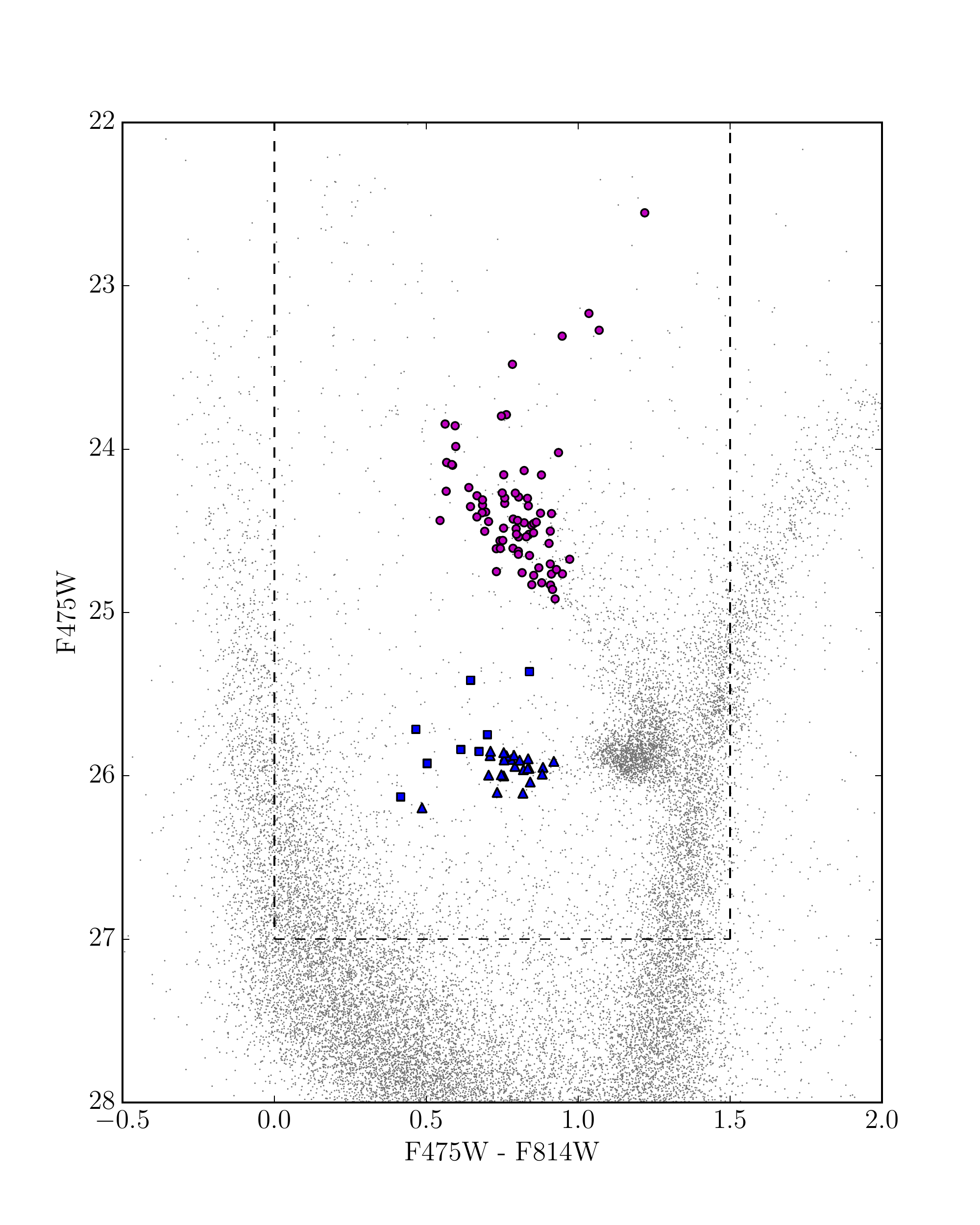}
\caption{\small{The CMD of DDO210 focusing on the pulsating variable
    stars. The grey points indicate the entire photometric sample of
    stars, while the coloured shapes indicate the pulsating
    variables. Each symbol represents a Cepheid (magenta
    circles), an RRab star (blue triangles), or an RRc star (blue
    squares). Phase-averaged mean magnitudes and colours calculated
    from the best-fitting template light curve are used to plot the
    variables. The dashed line shows the selection region adopted to
    find potential pulsating stars.\label{fig:var_cmd}}}
\end{figure}

\begin{figure*}
\includegraphics[width=2\columnwidth]{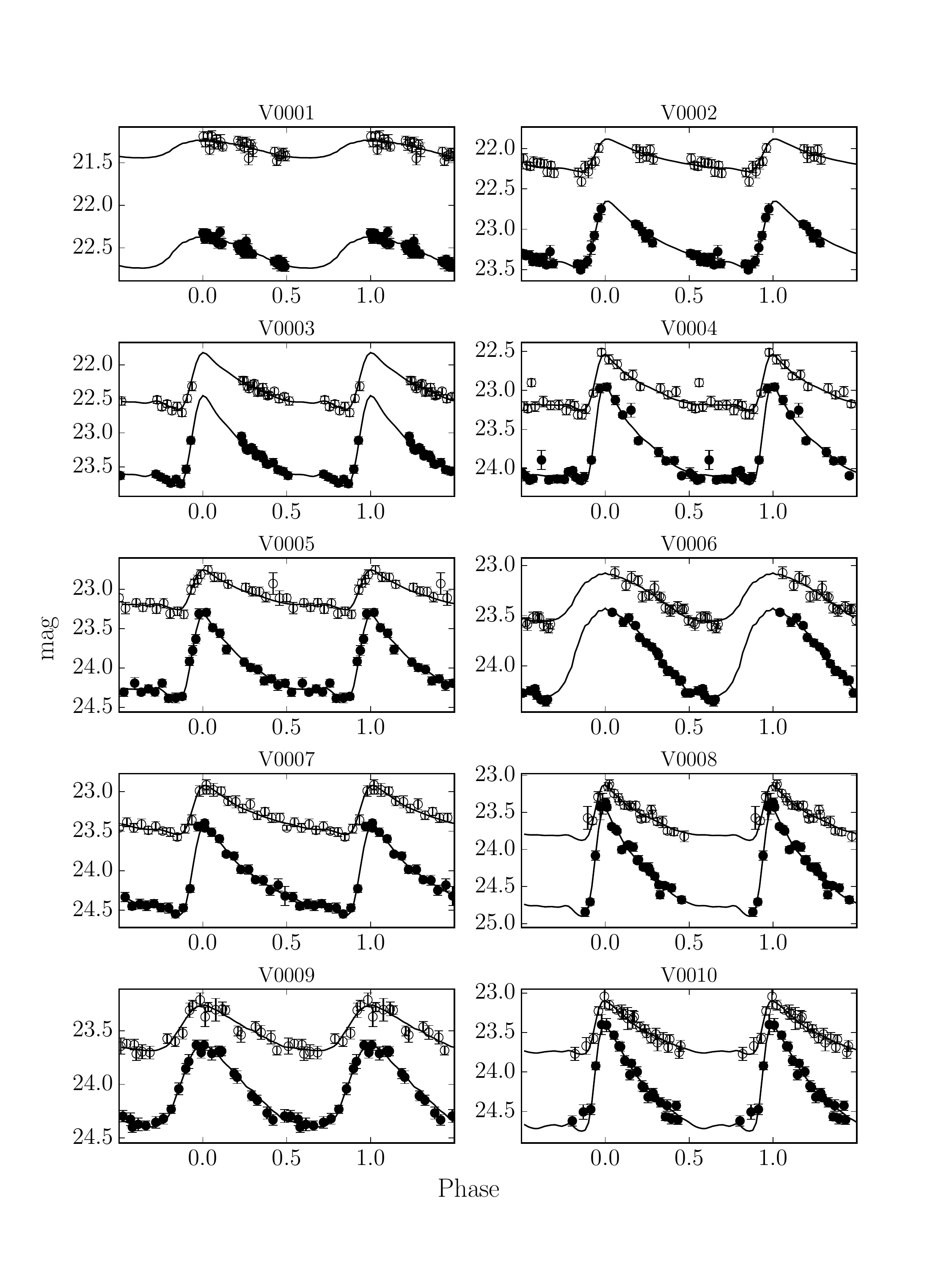}
\caption{\small{Example light curves of some Cepheid variables in
    DDO210. Open circles represent F814W observations, while filled
    circles are F475W observations. The solid lines illustrate the
    best-fitting templates to each light curve.\label{fig:clc}}}
\end{figure*}

\begin{figure*}
\includegraphics[width=2\columnwidth]{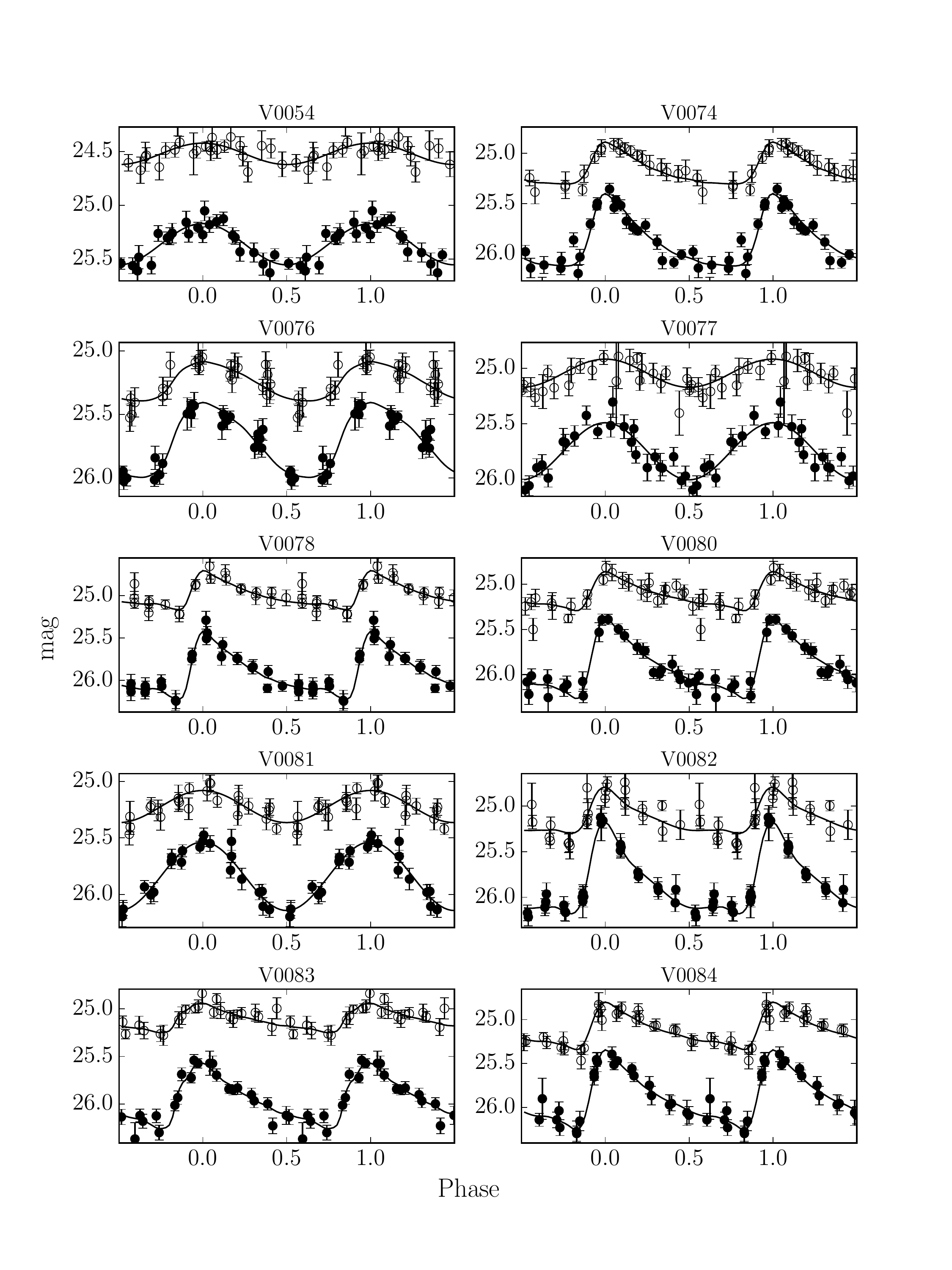}
\caption{\small{Same as Fig. \ref{fig:clc} for some RR Lyrae variables.\label{fig:rlc}}}
\end{figure*}

Stars which
failed to pass this vetting procedure with reasonable properties were
removed from the sample. This left 107 (75 Cepheid; 32 RR Lyrae) high confidence variable stars
in our sample. Manual
fitting with \texttt{FITLC} was required for 4 RR Lyrae stars and 7 Cepheids, accounting
for 10\% of our total pulsating variable sample. The CMD locations of our
final variable star sample are shown in Fig. \ref{fig:var_cmd}. Some example
light curves of each of these variables, folded with the best-fitting
period, are plotted in Figs. \ref{fig:clc} and
\ref{fig:rlc}. Properties of each of these candidate pulsators are
presented in Table \ref{tbl:varprops}.

\begin{table*}
\caption{Properties of the pulsating variables in
  DDO210.\label{tbl:varprops}}
\scalebox{0.6}{
\begin{tabular}{cllccccccc}
\hline
Star ID & RA (J2000) & Dec (J2000) &
Type & Period (days) & A$_B$
    (mag) & A$_I$
    (mag) & $\langle B \rangle$ (mag) & $\langle I
    \rangle$ (mag) & $\langle B - I \rangle$ (mag)\\
\hline
V0054 & 20 46 48.58750 & -12 -51 -21.23784 & RRc & 0.342 & 0.390 & 0.200 & 25.492 & 24.509 & 0.983\\
V0074 & 20 46 47.22113 & -12 -51 -35.63064 & RRab & 0.620 & 0.769 & 0.418 & 25.994 & 25.147 & 0.847\\
V0076 & 20 46 51.51648 & -12 -50 -52.51020 & RRc & 0.330 & 0.645 & 0.308 & 25.793 & 25.234 & 0.559\\
V0077 & 20 46 50.33340 & -12 -51 -10.18836 & RRc & 0.450 & 0.556 & 0.255 & 25.869 & 25.031 & 0.837\\
V0078 & 20 46 47.77346 & -12 -50 -51.88416 & RRab & 0.732 & 0.873 & 0.460 & 26.056 & 24.972 & 1.083\\
V0080 & 20 46 48.79860 & -12 -50 -30.80076 & RRab & 0.635 & 0.944 & 0.431 & 26.015 & 25.100 & 0.915\\
V0081 & 20 46 45.52898 & -12 -50 -58.67916 & RRc & 0.376 & 0.662 & 0.284 & 25.944 & 25.210 & 0.734\\
V0082 & 20 46 47.98013 & -12 -50 -28.58712 & RRab & 0.597 & 1.049 & 0.505 & 25.963 & 25.119 & 0.844\\
V0083 & 20 46 47.03861 & -12 -50 -56.49756 & RRab & 0.639 & 0.752 & 0.309 & 26.084 & 25.098 & 0.986\\
V0084 & 20 46 50.66448 & -12 -50 -52.12644 & RRab & 0.595 & 0.940 & 0.538 & 25.999 & 25.104 & 0.895\\
V0086 & 20 46 49.49820 & -12 -51 -6.60276 & RRab & 0.561 & 0.737 & 0.406 & 25.992 & 25.067 & 0.925\\
V0087 & 20 46 52.94460 & -12 -51 -8.88444 & RRab & 0.650 & 0.667 & 0.302 & 26.100 & 25.067 & 1.032\\
V0088 & 20 46 50.28744 & -12 -51 -5.53860 & RRab & 0.629 & 0.785 & 0.406 & 25.994 & 25.094 & 0.901\\
V0089 & 20 46 51.78494 & -12 -50 -42.38304 & RRc & 0.374 & 0.509 & 0.278 & 25.967 & 25.162 & 0.805\\
V0090 & 20 46 49.32547 & -12 -50 -55.41936 & RRab & 0.617 & 0.994 & 0.472 & 26.060 & 25.130 & 0.930\\
V0091 & 20 46 49.43150 & -12 -51 -28.76688 & RRab & 0.600 & 0.716 & 0.327 & 26.021 & 25.078 & 0.943\\
V0092 & 20 46 44.41512 & -12 -50 -33.30564 & RRab & 0.642 & 0.691 & 0.362 & 26.081 & 25.124 & 0.957\\
V0094 & 20 46 50.87971 & -12 -51 -48.50532 & RRab & 0.606 & 1.068 & 0.482 & 26.164 & 25.173 & 0.992\\
V0096 & 20 46 48.66934 & -12 -51 -22.93704 & RRab & 0.649 & 0.362 & 0.229 & 26.119 & 25.093 & 1.026\\
V0097 & 20 46 49.56259 & -12 -51 -1.17468 & RRab & 0.569 & 1.256 & 0.771 & 26.103 & 25.219 & 0.883\\
V0099 & 20 46 47.52367 & -12 -49 -53.79924 & RRab & 0.522 & 1.227 & 0.582 & 26.097 & 25.270 & 0.827\\
V0102 & 20 46 49.54975 & -12 -50 -54.75804 & RRab & 0.510 & 1.453 & 0.723 & 26.197 & 25.340 & 0.858\\
V0111 & 20 46 47.98416 & -12 -50 -31.82388 & RRc & 0.250 & 0.534 & 0.225 & 26.211 & 25.715 & 0.496\\
V1036 & 20 46 55.98754 & -12 -50 -57.17004 & RRc & 0.472 & 0.601 & 0.277 & 25.525 & 24.753 & 0.772\\
V1043 & 20 46 52.33248 & -12 -50 -23.35956 & RRab & 0.605 & 1.063 & 0.443 & 26.011 & 25.124 & 0.887\\
V1044 & 20 46 55.76921 & -12 -50 -44.24928 & RRab & 0.666 & 0.478 & 0.311 & 26.020 & 25.045 & 0.975\\
V1045 & 20 46 55.55026 & -12 -51 -14.49324 & RRab & 0.596 & 1.095 & 0.325 & 25.964 & 25.080 & 0.884\\
V1048 & 20 46 54.01258 & -12 -50 -58.06176 & RRab & 0.663 & 0.750 & 0.368 & 26.082 & 25.103 & 0.978\\
V1049 & 20 47 0.18739 & -12 -50 -42.15840 & RRab & 0.569 & 1.210 & 0.591 & 26.102 & 25.223 & 0.878\\
V1051 & 20 46 52.85431 & -12 -50 -37.57992 & RRab & 0.550 & 1.060 & 0.517 & 26.267 & 25.313 & 0.954\\
V1053 & 20 46 56.76238 & -12 -50 -26.97792 & RRc & 0.280 & 0.340 & 0.260 & 26.018 & 25.415 & 0.603\\
V1057 & 20 46 54.88075 & -12 -51 -6.19776 & RRab & 0.600 & 0.458 & 0.186 & 26.287 & 25.706 & 0.581\\
V0001 & 20 46 48.97092 & -12 -50 -53.82204 & Ceph & 5.150 & 0.442 & 0.214 & 22.827 & 21.332 & 1.495\\
V0002 & 20 46 55.35701 & -12 -51 -50.00616 & Ceph & 3.259 & 0.968 & 0.405 & 23.357 & 22.115 & 1.242\\
V0003 & 20 46 51.64591 & -12 -50 -54.38436 & Ceph & 2.213 & 1.452 & 0.835 & 23.462 & 22.335 & 1.128\\
V0004 & 20 46 53.18671 & -12 -51 -55.13688 & Ceph & 1.426 & 1.372 & 0.714 & 23.892 & 22.999 & 0.893\\
V0005 & 20 46 50.68507 & -12 -50 -37.68144 & Ceph & 1.631 & 1.219 & 0.507 & 24.178 & 23.063 & 1.115\\
V0006 & 20 46 45.28646 & -12 -50 -17.51496 & Ceph & 0.969 & 0.956 & 0.483 & 23.983 & 23.315 & 0.668\\
V0007 & 20 46 48.31193 & -12 -51 -52.58412 & Ceph & 1.526 & 1.225 & 0.623 & 24.253 & 23.286 & 0.968\\
V0008 & 20 46 43.86763 & -12 -50 -40.31628 & Ceph & 1.034 & 1.588 & 0.751 & 24.538 & 23.613 & 0.925\\
V0009 & 20 46 45.51108 & -12 -50 -12.33744 & Ceph & 0.677 & 0.830 & 0.423 & 24.175 & 23.496 & 0.679\\
V0010 & 20 46 51.82291 & -12 -51 -10.62108 & Ceph & 0.959 & 1.443 & 0.675 & 24.436 & 23.546 & 0.890\\
V0011 & 20 46 45.74947 & -12 -51 -11.95200 & Ceph & 0.685 & 1.063 & 0.500 & 24.191 & 23.490 & 0.701\\
V0012 & 20 46 50.71901 & -12 -50 -18.01356 & Ceph & 0.895 & 0.807 & 0.297 & 24.271 & 23.383 & 0.888\\
V0013 & 20 46 49.26427 & -12 -50 -13.07616 & Ceph & 0.955 & 1.417 & 0.724 & 24.388 & 23.592 & 0.795\\
V0014 & 20 46 51.50926 & -12 -51 -42.14664 & Ceph & 0.718 & 0.701 & 0.340 & 24.412 & 23.522 & 0.890\\
V0015 & 20 46 47.23687 & -12 -50 -6.20844 & Ceph & 0.998 & 1.267 & 0.565 & 24.595 & 23.594 & 1.001\\
V0016 & 20 46 50.88022 & -12 -50 -38.72940 & Ceph & 0.621 & 0.716 & 0.290 & 24.342 & 23.576 & 0.765\\
V0017 & 20 46 50.23925 & -12 -50 -40.32816 & Ceph & 1.192 & 1.295 & 0.625 & 24.543 & 23.459 & 1.083\\
V0018 & 20 46 50.30143 & -12 -50 -50.44524 & Ceph & 0.668 & 0.720 & 0.339 & 24.458 & 23.640 & 0.818\\
V0019 & 20 46 51.83731 & -12 -50 -48.18336 & Ceph & 1.240 & 1.538 & 0.761 & 24.495 & 23.556 & 0.939\\
V0020 & 20 46 51.61224 & -12 -50 -55.80672 & Ceph & 0.581 & 0.659 & 0.356 & 24.353 & 23.675 & 0.678\\
V0021 & 20 46 52.95365 & -12 -52 -12.38952 & Ceph & 0.684 & 0.918 & 0.450 & 24.422 & 23.605 & 0.817\\
V0022 & 20 46 50.48246 & -12 -51 -3.94992 & Ceph & 1.240 & 1.029 & 0.498 & 24.569 & 23.542 & 1.026\\
V0023 & 20 46 54.68083 & -12 -52 -0.27660 & Ceph & 0.729 & 0.826 & 0.351 & 24.567 & 23.723 & 0.844\\
V0024 & 20 46 48.06578 & -12 -49 -34.85784 & Ceph & 0.701 & 0.672 & 0.338 & 24.607 & 23.674 & 0.933\\
V0025 & 20 46 52.73578 & -12 -51 -15.08364 & Ceph & 1.021 & 0.739 & 0.364 & 24.645 & 23.664 & 0.981\\
V0026 & 20 46 44.19120 & -12 -50 -33.40968 & Ceph & 0.606 & 0.940 & 0.484 & 24.497 & 23.668 & 0.829\\
V0027 & 20 46 51.72058 & -12 -50 -40.72740 & Ceph & 0.568 & 0.681 & 0.406 & 24.504 & 23.687 & 0.817\\
V0028 & 20 46 49.50566 & -12 -51 -7.62624 & Ceph & 0.644 & 0.784 & 0.371 & 24.694 & 23.804 & 0.889\\
V0029 & 20 46 53.75340 & -12 -51 -27.60156 & Ceph & 0.914 & 1.644 & 0.770 & 24.706 & 23.849 & 0.858\\
V0030 & 20 46 47.99251 & -12 -50 -41.84232 & Ceph & 0.627 & 0.644 & 0.301 & 24.674 & 23.790 & 0.884\\
V0031 & 20 46 49.11158 & -12 -51 -1.41660 & Ceph & 1.010 & 0.875 & 0.435 & 24.747 & 23.688 & 1.059\\
V0032 & 20 46 53.02282 & -12 -51 -5.37408 & Ceph & 0.585 & 1.102 & 0.588 & 24.522 & 23.869 & 0.653\\
V0033 & 20 46 49.92425 & -12 -51 -9.03276 & Ceph & 0.930 & 1.291 & 0.629 & 24.696 & 23.700 & 0.996\\
V0034 & 20 46 49.45104 & -12 -51 -1.83240 & Ceph & 0.940 & 1.161 & 0.490 & 24.925 & 23.793 & 1.132\\
V0035 & 20 46 49.53163 & -12 -50 -50.46612 & Ceph & 0.591 & 0.517 & 0.247 & 24.731 & 23.807 & 0.924\\
V0036 & 20 46 49.10290 & -12 -50 -37.92984 & Ceph & 0.872 & 1.143 & 0.543 & 24.776 & 23.788 & 0.989\\
V0037 & 20 46 50.65714 & -12 -51 -18.41616 & Ceph & 0.613 & 0.506 & 0.310 & 24.734 & 23.848 & 0.886\\
V0038 & 20 46 50.52168 & -12 -50 -46.53708 & Ceph & 0.904 & 0.533 & 0.296 & 24.903 & 23.836 & 1.067\\
V0039 & 20 46 49.79030 & -12 -50 -23.87436 & Ceph & 0.904 & 1.060 & 0.534 & 24.740 & 23.799 & 0.941\\
V0040 & 20 46 50.26063 & -12 -50 -17.06856 & Ceph & 1.110 & 0.760 & 0.407 & 24.989 & 23.928 & 1.061\\
V0041 & 20 46 50.68666 & -12 -50 -50.26020 & Ceph & 1.023 & 1.157 & 0.514 & 24.877 & 23.917 & 0.960\\
V0042 & 20 46 48.62952 & -12 -50 -59.83476 & Ceph & 0.877 & 1.091 & 0.547 & 24.855 & 23.998 & 0.857\\
V0043 & 20 46 44.30563 & -12 -50 -57.25320 & Ceph & 0.982 & 0.812 & 0.537 & 24.836 & 23.775 & 1.061\\
V0044 & 20 46 51.88819 & -12 -50 -57.41664 & Ceph & 0.878 & 0.671 & 0.271 & 24.951 & 23.920 & 1.031\\
V0045 & 20 46 44.73055 & -12 -50 -29.71968 & Ceph & 0.705 & 1.186 & 0.569 & 24.959 & 23.960 & 0.999\\
V0046 & 20 46 52.61818 & -12 -51 -3.99636 & Ceph & 0.894 & 0.692 & 0.311 & 25.005 & 23.925 & 1.079\\
V1001 & 20 46 56.29128 & -12 -50 -38.03388 & Ceph & 2.710 & 0.975 & 0.440 & 23.499 & 22.220 & 1.279\\
V1002 & 20 46 59.11416 & -12 -51 -20.16036 & Ceph & 1.810 & 1.482 & 0.747 & 23.591 & 22.668 & 0.922\\
V1003 & 20 46 57.23064 & -12 -51 -42.49656 & Ceph & 0.651 & 0.407 & 0.202 & 23.926 & 23.035 & 0.890\\
V1004 & 20 46 54.93190 & -12 -50 -51.88236 & Ceph & 0.775 & 0.546 & 0.264 & 23.959 & 23.246 & 0.713\\
V1005 & 20 46 52.35540 & -12 -50 -3.36192 & Ceph & 0.784 & 0.973 & 0.508 & 24.080 & 23.366 & 0.714\\
V1006 & 20 46 56.79665 & -12 -51 -17.93808 & Ceph & 0.645 & 0.714 & 0.352 & 24.191 & 23.492 & 0.700\\
V1007 & 20 46 56.47327 & -12 -51 -18.07668 & Ceph & 1.001 & 1.417 & 0.660 & 24.385 & 23.450 & 0.935\\
V1008 & 20 46 56.14488 & -12 -50 -41.49996 & Ceph & 1.280 & 0.966 & 0.460 & 24.330 & 23.300 & 1.029\\
V1009 & 20 46 54.45374 & -12 -51 -6.73524 & Ceph & 1.103 & 1.099 & 0.474 & 24.651 & 23.572 & 1.079\\
V1010 & 20 46 55.78406 & -12 -50 -31.56396 & Ceph & 1.018 & 1.293 & 0.579 & 24.570 & 23.604 & 0.967\\
V1011 & 20 46 49.64760 & -12 -49 -31.83312 & Ceph & 1.181 & 1.065 & 0.556 & 24.471 & 23.488 & 0.982\\
V1012 & 20 46 55.29806 & -12 -51 -3.96792 & Ceph & 0.965 & 1.471 & 0.775 & 24.447 & 23.678 & 0.769\\
V1013 & 20 46 54.07574 & -12 -51 -13.27680 & Ceph & 0.480 & 1.439 & 0.698 & 24.449 & 23.576 & 0.873\\
V1014 & 20 46 53.67084 & -12 -50 -58.58376 & Ceph & 1.080 & 1.326 & 0.676 & 24.500 & 23.526 & 0.974\\
V1015 & 20 46 53.71486 & -12 -50 -57.96780 & Ceph & 1.036 & 1.417 & 0.548 & 24.591 & 23.575 & 1.016\\
V1016 & 20 46 56.17759 & -12 -50 -29.73336 & Ceph & 0.993 & 0.657 & 0.368 & 24.655 & 23.688 & 0.966\\
V1017 & 20 46 54.85265 & -12 -50 -29.90580 & Ceph & 0.756 & 0.568 & 0.263 & 24.558 & 23.619 & 0.938\\
V1018 & 20 46 51.97082 & -12 -49 -43.11660 & Ceph & 0.939 & 1.050 & 0.589 & 24.650 & 23.710 & 0.940\\
V1019 & 20 46 52.79429 & -12 -50 -40.07328 & Ceph & 0.670 & 0.578 & 0.265 & 24.601 & 23.713 & 0.888\\
V1020 & 20 46 55.07467 & -12 -50 -50.70444 & Ceph & 0.602 & 0.686 & 0.313 & 24.527 & 23.731 & 0.796\\
V1021 & 20 46 56.47325 & -12 -51 -2.04552 & Ceph & 0.702 & 0.461 & 0.260 & 24.670 & 23.717 & 0.953\\
V1022 & 20 46 52.33610 & -12 -50 -36.21120 & Ceph & 1.020 & 1.308 & 0.548 & 24.652 & 23.638 & 1.013\\
V1023 & 20 46 56.23015 & -12 -50 -47.39568 & Ceph & 1.040 & 1.310 & 0.367 & 24.903 & 23.744 & 1.159\\
V1024 & 20 46 55.20768 & -12 -51 -6.00696 & Ceph & 0.632 & 1.124 & 0.518 & 24.625 & 23.794 & 0.831\\
V1025 & 20 46 58.72258 & -12 -51 -32.84100 & Ceph & 0.935 & 1.004 & 0.536 & 24.883 & 23.787 & 1.096\\
V1026 & 20 46 54.81619 & -12 -51 -13.88988 & Ceph & 0.864 & 0.724 & 0.315 & 24.761 & 23.821 & 0.940\\
V1027 & 20 46 56.39974 & -12 -49 -47.28360 & Ceph & 0.899 & 0.852 & 0.409 & 24.902 & 23.899 & 1.003\\
V1028 & 20 46 53.69839 & -12 -50 -57.49008 & Ceph & 0.928 & 0.829 & 0.297 & 24.864 & 23.836 & 1.028\\
V1030 & 20 46 50.64456 & -12 -49 -54.85404 & Ceph & 0.870 & 0.649 & 0.314 & 25.079 & 23.997 & 1.082\\
\hline
\end{tabular}}
\end{table*}

\subsection{Simulations}
\label{sub:sim}
We performed artificial light curve
simulations to assess the accuracy and precision of our automated fitting
procedure on this data set (see Ordo\~{n}ez, Yang, \& Sarajedini (2014) and references
therein). This involved using the Fourier templates from \texttt{RRFIT}
to create artificial light curves. The properties of these artificial
light curves were chosen to best represent the types of pulsating variables
found in DDO210. We sampled the appropriate period, amplitude,
magnitude, and colour ranges in which the RR Lyrae and Cepheids were
found to lie. Each of these properties was sampled from a uniform
distribution and assigned to each individual light curve. For the
artificial RR Lyrae light curves, the amplitude constraint in Equation
(\ref{eq:ampconstvgm}) was applied. Observations
on these artificial F475W,F814W light curves were then simulated utilizing the
cadence of our dataset and realistic photometric errors. 

Unlike the other studies employing these artificial
light curve simulations, we have accounted for ranges in magnitude and
colour, utilizing the photometric errors as a function of magnitude in
each the F475W and F814W bands. We modeled the photometric error
function as a constant value for bright magnitudes, and an exponential
function for fainter magnitudes. This model was then fit to the errors from each individual
photometric measurement for every star in our sample. In this way, we
were able to realistically assess how our light curve analysis
accuracy and completeness vary as a function of magnitude. 

\begin{figure*}
\includegraphics[width=2\columnwidth]{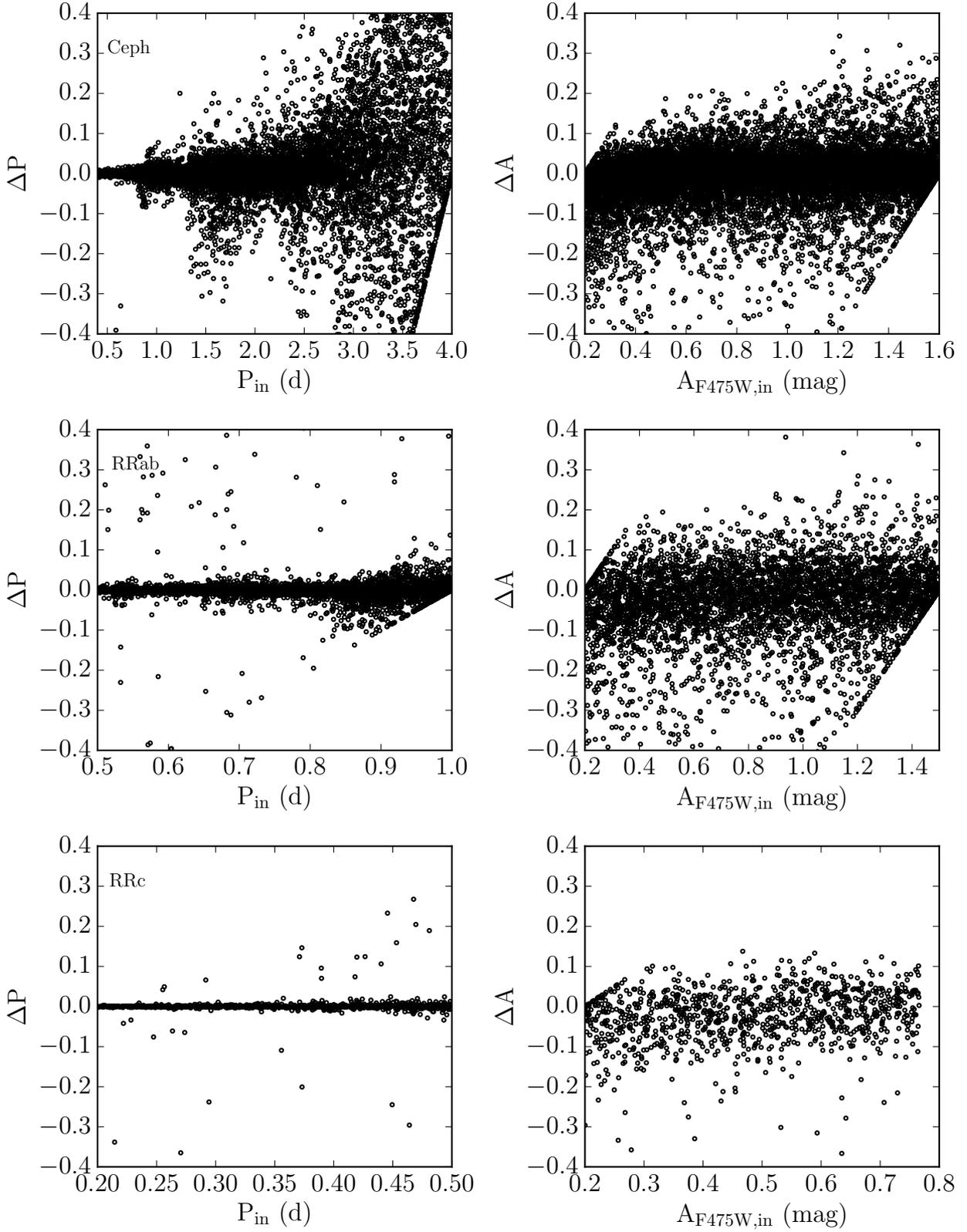}
\caption{\small{Results of the light curve simulations. The left
    panels show the period error as a function of input period, while
    the right panels show the amplitude error versus input F475W amplitude.\label{fig:simres}}}
\end{figure*}

Once the simulated observations were generated, we then input these
into \texttt{RRFIT} exactly as we did for our real light curves. The
difference between the synthetic input and best-fitting light curve
parameters for each artificial light curve then gives
estimates of the uncertainties inherent in the fitting procedure. We
show the results of these simulations for the Cepheid, RRab, and RRc
stars in Fig. \ref{fig:simres}. 

Given that these data cover a baseline of $\sim$2.5 days, it is
expected that periods greater than this to be recovered with
significantly less accuracy than for shorter periods. This is clearly
visible in the Cepheid simulations. However, for periods shorter than
2.5 days, the period errors remain constant and reasonably
small. Only 3 of the 75 Cepheid candidates were fit with periods
longer than 2.5 days, and the longest period Cepheid (5.15 days) was
fit manually in FITLC since it fell outside of the explored period
range. Therefore, we expect the observing baseline to affect the
periods of only 2
out of the 107 variables in our sample. 

To gauge the quantitative uncertainties inherent in the fitting
routine from these simulations, we perform the following statistical
test discussed in \citet{ord14}. From our sample of artificial light
curves, we sample the corresponding number of identified stars (75 Cepheid;
24 RRab: 8 RRc) and calculate the average error in the light curve parameters
($\Delta$P, $\Delta$A$_{\mathrm{F475W}}$, etc.). This sampling is
repeated 10,000 times to build a statistical distribution of the
expected errors. These distributions then represent good estimators
for the systematic and random errors inherent in the fitting
procedure. Specifically, the peak of the distribution reveals any
systematic errors, and the spread estimates the random uncertainties
(standard error) in each parameter. Gaussian fits to the final error distributions
therefore yield our fiducial uncertainty values. These uncertainties
are presented in Table \ref{tbl:err} for each type of simulated
variable. Since only 2 Cepheids will be affected by the large
uncertainties at P$>$2.5 days, we restricted this test to artificial
light curves with periods less than this value. 
\begin{table*}
\caption{Uncertainties estimated from the light curve
  simulations.\label{tbl:err}}
\scalebox{0.75}{
\begin{tabular}{ccccccc}
\hline
Variable & $\sigma_{P,systematic}$ (d) &
$\sigma_{P,random}$ (d) &
$\sigma_{AF475W,systematic}$ (mag) &
$\sigma_{AF475W,random}$ (mag) &
$\sigma_{mF475W,systematic}$ (mag) &
$\sigma_{mF475W,random}$ (mag)\\
\hline
Cepheid & -0.0044 & 0.0058 & -0.0233 & 0.0103 & -0.0004 & 0.0043\\
RRab & -0.0011 & 0.0042 & -0.0399 & 0.0257 & 0.0049 & 0.0103\\
RRc & 0.0007 & 0.0026 & -0.0213 & 0.0228 & -0.0022 & 0.0073\\
\hline
\end{tabular}}
\end{table*}

We conclude this section by pointing out that for most of the
simulated light curve parameters, the systematic errors are smaller
than the random errors. However, the simulations reveal that we may be
underestimating the amplitudes for the Cepheid and RRab light curves
by 0.02 and 0.04 mag, respectively. We tested to see if this affected
the metallicity estimates for the RRab stars (see Section
\ref{sub:rrab_met}) by subtracting this offset from the amplitudes,
and it turned out to drive the metallicity estimates higher by
$\sim$0.05 dex. Considering that this is smaller than the other
sources of uncertainty for this calculation, we do not expect these
errors to impact the main results of this paper significantly.

\section{RR Lyrae stars}
\label{sec:rrl}
The relative paucity of RR Lyrae stars in DDO210, especially
compared with the younger Cepheids, is fully consistent with the
appearance of its CMD. The red HB morphology and strong red clump (RC) in this
galaxy indiciate a younger average age for the stellar population. The
small number of RR Lyrae stars implies a weaker star formation rate at
times $\ga$10 Gyr ago, which is fully consistent with the
appearance of the other features. This agrees well with the
synthetic CMD, SFH analysis of \citet{col14}. Their Fig. 4 shows a
distinct minimum just before a lookback time of 10 Gyr. 

\subsection{Metallicity of RRab stars}
\label{sub:rrab_met}
In addition to providing insights into the SFH of a galaxy, the RR
Lyrae stars also inform us regarding the chemical enrichment of their
host system through their metallicity distribution function
(MDF). As discussed in Section \ref{sec:intro}, many studies have
provided methods of calculating an iron abundance, [Fe/H], from the
shape of the light curves of RR Lyrae stars. Considering that the
dataset employed in this study is not well-suited for a Fourier
analysis on each light curve, we utilized the relation of
\citet{alc00} to calculate [Fe/H] for each individual RRab star:
\begin{equation} \label{eq:met}
\mathrm{[Fe/H]} = -8.85[log P_{ab} + 0.15A_V] - 2.60
\end{equation}
Since we did not have $V$ band imaging for these stars, we opted to
estimate $A_V$ for each RRab star using the relation between $A_B$
and $A_I$ from \citet{df99}. 

\begin{figure}
\includegraphics[width=\columnwidth]{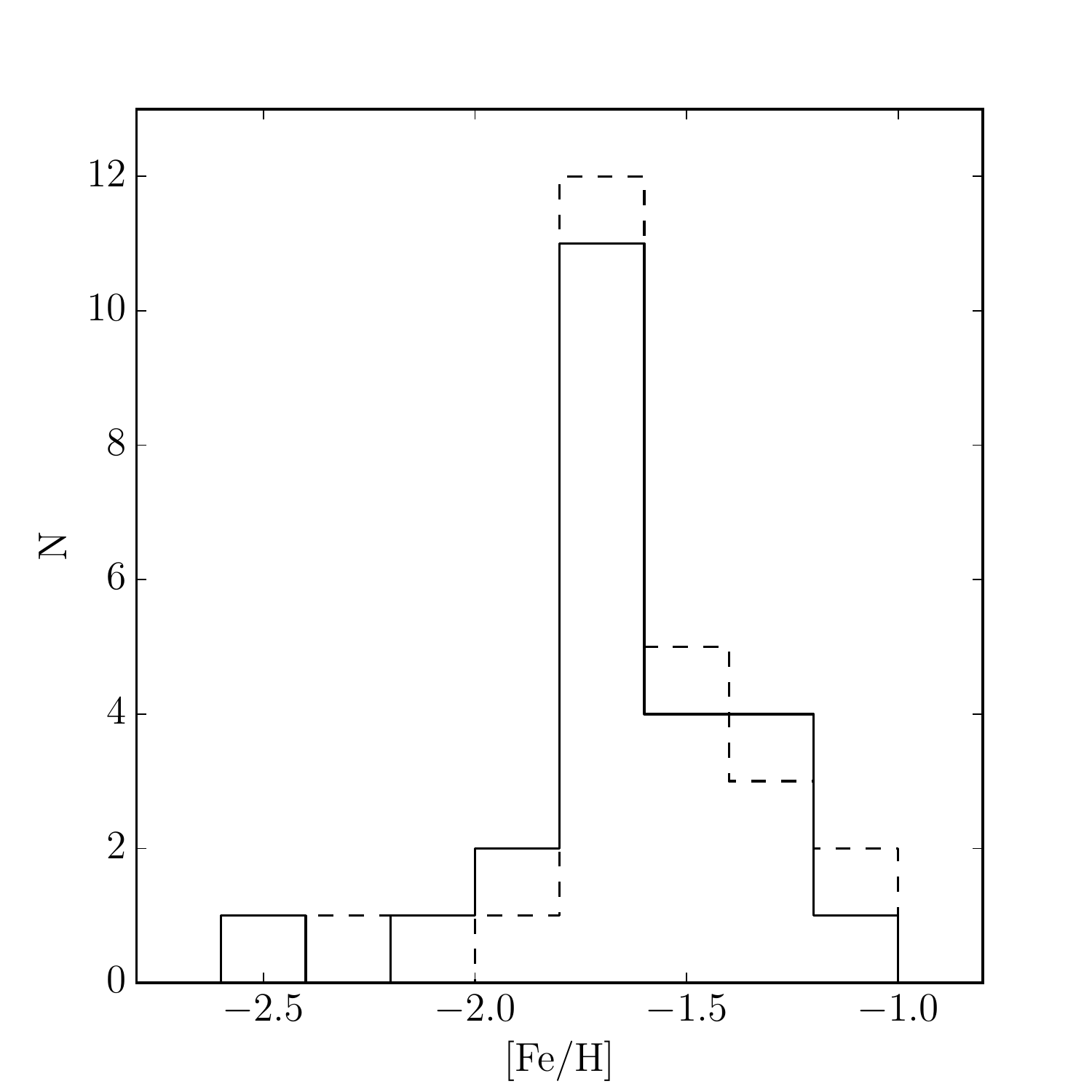}
\caption{\small{The MDF of the RRab stars using the relation of
    \citet{alc00} to calculate the metallicities. The solid line
    represents the MDF calculated converting $A_I$ to $A_V$. The
    dashed line is the same but estimating $A_V$ from $A_B$.\label{fig:rrab_mdf}}}
\end{figure}
Fig. \ref{fig:rrab_mdf} shows the RRab MDF calculated using both the
$I$-band and $B$-band amplitudes. The MDFs show no significant
difference, and the mean metallicity in each case is
$\langle$[Fe/H]$\rangle = -1.65\pm 0.11$ dex using $A_I$ and
$\langle$[Fe/H]$\rangle = -1.61\pm 0.11$ dex using $A_B$, both in full
agreement. We thus take a singular estimate by averaging these two
values together to obtain $\langle$[Fe/H]$\rangle = -1.63\pm 0.11$
dex. Assuming the
effects of $\alpha$-enhancement are small, this also agrees with the
AMR from the synthetic CMD analysis of \citet{col14}. Their Fig. 5
shows this AMR, and at a lookback time of approximately 11 Gyr, this
average metallicity is [M/H] $\sim -1.7$ dex. The RRab stars, which
formed at approximately this time, support the results of the analysis
of \citet{col14}. 

\subsection{Bailey diagram}
\label{sub:bailey}
\begin{figure}
\includegraphics[width=\columnwidth]{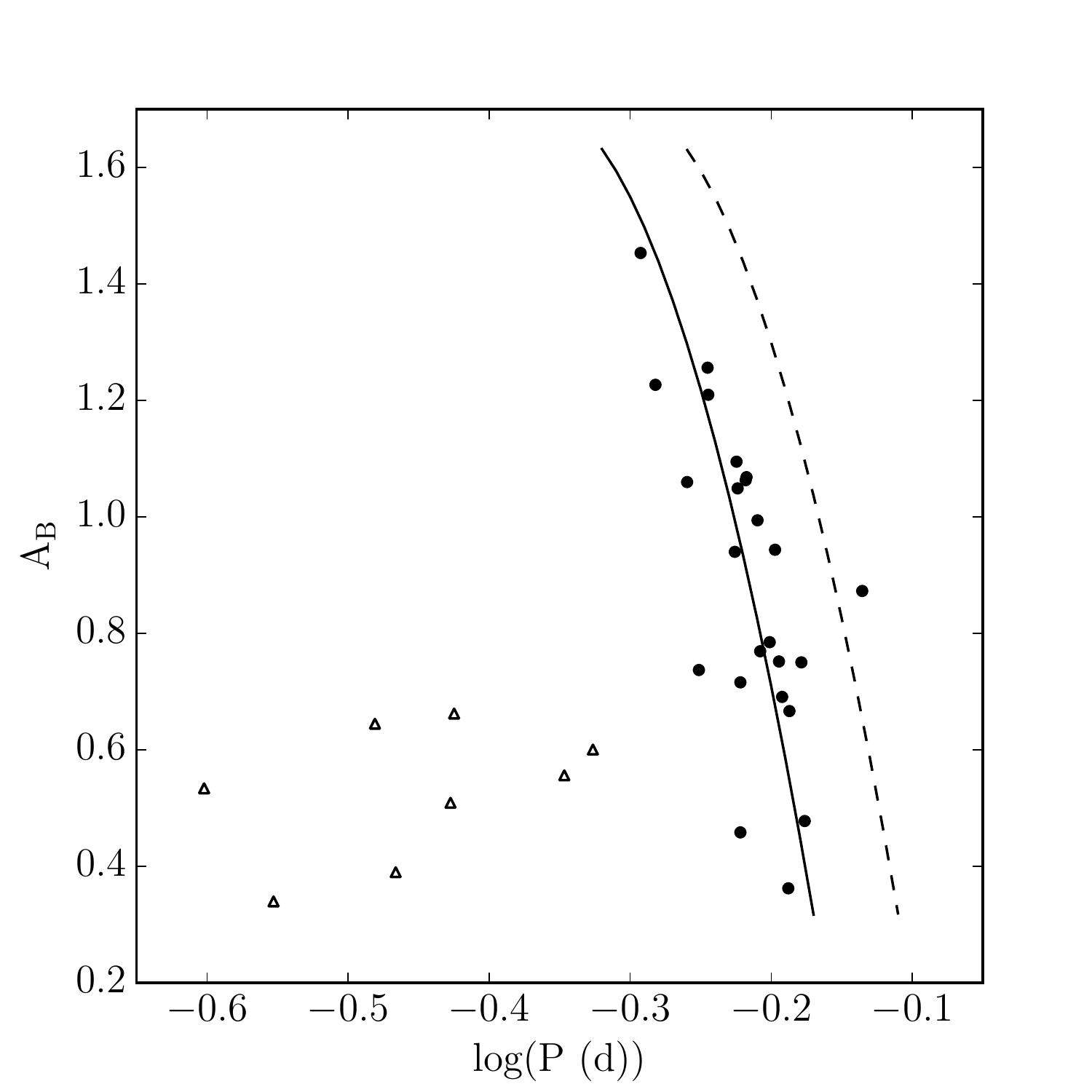}
\caption{\small{The Bailey diagram for the RR Lyrae stars in DDO210. RRab
  stars are represented as filled circles, while the open triangle
  represent the RRc stars. Plotted as the solid and dashed lines are
  the Oosterhoff I and II trend lines from \citet{cac05}.\label{fig:bailey}}}
\end{figure}
An important diagnostic of an RR Lyrae population is the Bailey
(period-amplitude) diagram. We show this for the 32 RR Lyrae stars
within DDO210 in Fig. \ref{fig:bailey}. Also plotted in this figure
are the loci for the two different Oosterhoff populations in the
Galactic globular clusters (GGCs) from Cacciari, Corwin, \& Carney (2005). While most of the RRab
stars lie around the Oosterhoff type I (OoI) locus, their mean period
$\langle$P$\rangle = 0.609\pm 0.011$ d and mean metallicity
$\langle$[Fe/H]$\rangle = -1.63\pm 0.11$ dex place this galaxy in the
so-called Oosterhoff gap, along with the majority of dSph
galaxies. Additionally, the first-overtone to fundamental mode
population ratio, (RRc)/(RRc + RRab) = 0.25, is consistent with an
Oosterhoff-intermediate classification for this galaxy.

The discrepancy between the appearance of the Bailey diagram and the
other Oosterhoff classification methods have been noted in previous
studies \citep{ord14, ste14}. \citet{ste14, fio15} discuss this issue
in detail, and conclude that the absence of high-amplitude,
short-period (HASP) RRab stars in dSph and
ultra-faint dwarf (UFD) galaxies present a
fundamental difference between the ancient RR Lyrae population in
dSph/UFDs and those in the GGCs and Galactic Halo. This HASP absence,
which \citet{fio15} attribute to the metal-poor environments forming
the RR Lyrae stars, skews the mean RRab period to longer
values. Therefore, it seems that DDO210 hosts an RR Lyrae population
similar to the dSph and UFDs in the LG owing to their similarly
metal-poor environments at early ages.

\subsection{Distance to RRab stars}
\label{sub:rrab_mu}
The distance to DDO210 has been a matter of some debate since its
discovery. Most recent distance estimates to this galaxy using the
TRGB place DDO210 at a distance modulus of $\umu \sim 25$ \citep{mcc05,
  jac09}, and the most recent synthetic CMD analysis by \citet{col14}
report a distance modulus of $\umu = 24.95\pm 0.10$. We now compute the
distance to DDO210 using the period-luminosity-metallicity (PLZ) relation
from \citet{cat04}:
\begin{equation}
M_I = 0.471 -1.132 \mathrm{log} P + 0.205 \mathrm{log} Z
\end{equation}
Using the RRab metallicities from Section \ref{sub:rrab_met}
calculated with the $I$-band amplitudes, we
convert to log$Z$ using the following relation: $\mathrm{log} Z =
[Fe/H] - 1.765$ to remain consistent with \citet{cat04}. Again, here
we have not corrected for the effects of $\alpha$-enhancement in the
absence of a strong constraint for [$\alpha$/Fe] for these
stars. Apparent mean magnitudes of these RRab stars have been
corrected for extinction using $A_I = 0.076$ mag from \citet{sf11} for DDO210
retrieved from NED\footnote{\url{https://ned.ipac.caltech.edu/}}. 

The mean distance modulus of the RRab stars calculated in this way is
$\umu = 25.07\pm 0.12$ where the error bar represents one standard
deviation. We note that the effects of different helium abundances on
this RR Lyrae PLZ relation have not been fully explored, and
significant helium abundance in the DDO210 RR Lyrae stars could
present a significant systematic error in this distance
determination. There may also be some systematic error resulting from
$\alpha$-enhancement when calculating log$Z$. Even considering these
uncertainties, our RR Lyrae distance is roughly consistent with other
distance determinations to this galaxy within the uncertainties. The
RR Lyrae distance does however seem to lie on the farther end of the
distribution of previous distance measurements. In fact, our distance
is closer to that of \citet{mcc05} ($\umu = 25.15\pm 0.08$).

\subsection{Peculiar RR Lyrae candidates}
\label{sub:arrl}
We now turn to discussing a few anomalous RR Lyrae candidates within
our sample. The CMD of DDO210 in Fig. \ref{fig:var_cmd} shows two
bright outliers in the RR Lyrae population near
 $m_{\mathrm{F475W}}=25.4$ mag and colours of (F475W - F814W) $=0.6$
 and (F475W - F814W) $=0.8$ mag. Neither of these RRc stars appear as
outliers in the Bailey diagram, nor do they display anomalous light
curves. It is evident from Fig. \ref{fig:var_cmd} that most of the RR Lyrae
candidates that we identified occupy the expected region in the
CMD, except for the two aforementioned bright outliers
lie clearly brighter the HB. In fact,
these stars appear to lie in between the bulk RR Lyrae population and the
Cepheids. It is possible that these stars are faint ACs, but this
seems unlikely given their separation from the bulk Cepheid population
in the CMD and PL relations (see Section \ref{sec:ceph} for a full
discussion of the Cepheids). We also considered the possibility that
these stars are type II Cepheids (BL Her) stars, but ruled this out
because BL Her stars generally have periods between 1 and 4 days (
these outliers both have periods near 0.4 days). Another
possibility is that these stars contain unresolved companions, making
them appear brighter. This is more likely the case as Cepheids with
such short periods are uncommon. 

One more peculiar RR Lyrae candidate deserves discussion. Immediately
apparent upon examining Fig.
\ref{fig:bailey} is the one outlying RRab star near the OoII
locus. The longest period RR Lyrae in our sample, this star is also
the most metal-poor RRab star. While it is tempting to discard this
star as a contaminating Type II or anomalous Cepheid, we note that
this star lies right on the HB ($m_{\mathrm{F475W}}=25.91$ mag) in the CMD near the red end of the RR
Lyrae gap. It may therefore be a very metal-poor RR Lyrae star formed within
DDO210. Considering once again the analysis of \citet{col14} for
comparison, we see that the AMR for the galaxy has the largest
metallicity dispersion at a lookback time of 11 Gyr, corresponding to
the age of the RR Lyrae stars. In fact, the rms metallicity range for
that time bin extends down to metallicities of [M/H] $ = -2.5$
dex. Therefore, such a metal-poor RR Lyrae star is not inconsistent
with the synthetic CMD analysis of \citet{col14}. 

\section{Cepheids: anomalous or classical?}
\label{sec:ceph}
\subsection{Comparison with evolutionary tracks}
\label{sub:evo} 
The Cepheid variables dominate the pulsators found in DDO210.
However, as other authors have pointed out \citep{fio12}, this
region of the CMD hosting these stars is degenerate. That is to say that stellar
populations with different masses and metallicites can occupy this
region, including the Cepheid instability strip. In low-metallicity
environments, this makes distinction between these dfferent
populations difficult. We note that no Cepheids with periods longer
than $\sim$5 days were identified in our data. Given the quality of
the data at these bright magnitudes, we expect to flag variability at
these longer periods even with incomplete phase coverage. Therefore,
this absence of longer period Cepheids is more likely a reflection of
the Cepheid population itself rather than an observational bias.

To further complicate matters, recent investigations of the Cepheid
populations in dwarf galaxies have revealed that the distinction
between ACs and CCs is not as clear as once
thought. These studies have found that the period and luminosity distributions of
these two types of variables can overlap, making a separation based on
this diagram alone difficult \citep{gal04, fio12, cle12, ber13,
  ste14}. One other way to potentially distinguish between a
population of CCs and ACs is to utilize our theoretical understanding
of ACs. Since ACs are stars with masses less than the
transition mass between partial-degenerate, central He burning and
quiescent central He burning, one can compare the positions of stellar
evolutionary tracks near this mass ($\sim$2.1 M$_{\sun}$ for
[Fe/H]$\le$0.7 dex) with the observed Cepheid population
\citep{fio12}. 

\begin{figure*}
\includegraphics[width=2\columnwidth]{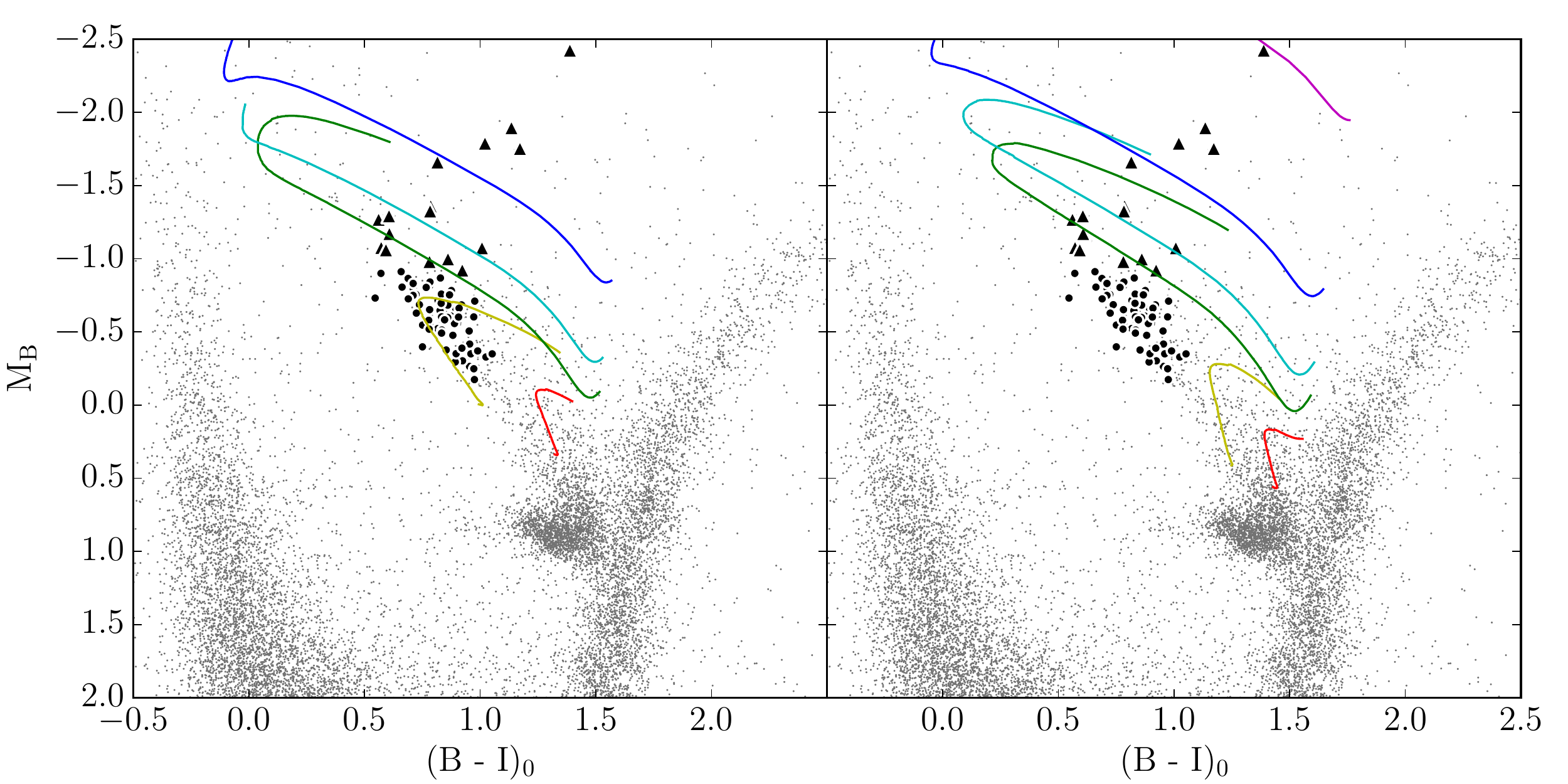}
\caption{\small{\emph{Left:} The CMD of DDO210 comparing Cepheid
    variables (black circles and triangles) with BaSTI He-burning, stellar
    evolutionary tracks (red to magenta: 1.8, 2.1, 2.8, 3.0, 3.5, and 5.0
    M$_{\sun}$) for [Fe/H] $ = -1.5$ dex. The triangles show Cepheids that
    are very likely CCs, while the filled circles show Cepheids that
    may be either ACs or CCs. \emph{Right:} Same as the
    left panel but with [Fe/H] $ = -1.3$ dex evolutionary tracks of
    the same masses.\label{fig:ceph_cmd}}}
\end{figure*}

We have performed this analysis for the Cepheids in DDO210, and the
results are shown in Fig. \ref{fig:ceph_cmd}. This shows the CMD of
DDO210 highlighting the Cepheid variables. The magnitudes have been
converted to absolute magnitudes using the RR Lyrae distance modulus
from Section \ref{sub:rrab_mu}, and colours de-reddened using the
reddening to DDO210 from \citet{sf11}. To attempt to deduce the nature
of the Cepheids in DDO210, we have
plotted the He-burning stages of the canonical, scaled-solar evolutionary tracks for stars of
different masses (1.8, 2.1, 2.8, 3.0, 3.5, and 5.0 M$_{\sun}$;
coloured lines in \ref{fig:ceph_cmd})  and
two different metallicities ([Fe/H] $ = -1.5$ dex and [Fe/H] $ = -1.3$
dex; left and right panels in Fig. \ref{fig:ceph_cmd},
respectively) from the BaSTI stellar evolutionary library \citep{pie04}. We did not explore
lower metallicities because such stars are deemed unlikely to have
been formed within the past 6 Gyr considering the AMR from
\citet{col14}. On the other hand, ACs are not expected to reach
metallicities above [Fe/H] $ = -1.3$ dex because they do not enter the
instability strip above this metallicity, so higher metallicity tracks
were not explored either.

From Fig. \ref{fig:ceph_cmd}, it is evident that the brighter
Cepheids in this galaxy are clearly CCs, irrespective of the
metallicity adopted for these stars. We have identified these stars as
lying near or brighter than the 2.8 M$_{\sun}$ track in both cases
(the triangles in Fig. \ref{fig:ceph_cmd}). The remaining Cepheids
cluster around the 2.1 M$_{\sun}$ track for the lower metallicity
case but not the higher metallicity case. Thus, we are presented with
a situation similar to what \citet{fio12} found in Leo I. That is, the
nature of the Cepheids in DDO210 seems to depend on the metallicity of
these stars. If they are sufficiently metal-poor, which \citet{kir13}
indicate is possible calculating $\langle$[Fe/H]$\rangle = -1.44$ dex
for DDO210, it is likely that
many of the fainter Cepheids are ACs. On the other hand, if these
stars are all characterized by a slightly higher metallicity, as is
suggested by the AMR of \citet{col14} for ages $\le$6 Gyr, then these
Cepheids should all be CCs. 

\subsection{PL relations}
\label{sub:pl}
\begin{figure*}
  $\begin{array}{cc}
    \includegraphics[width=\columnwidth]{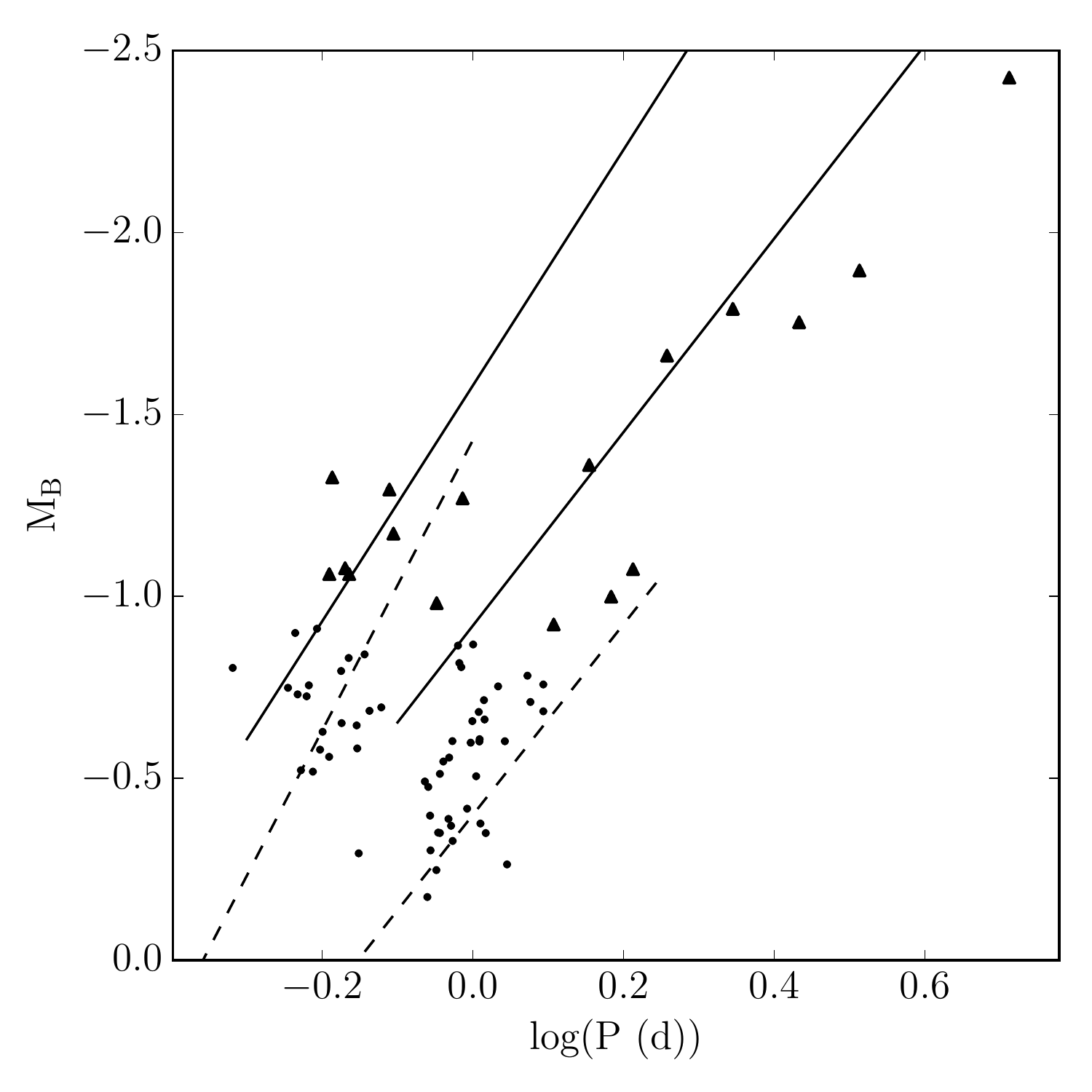}&
    \includegraphics[width=\columnwidth]{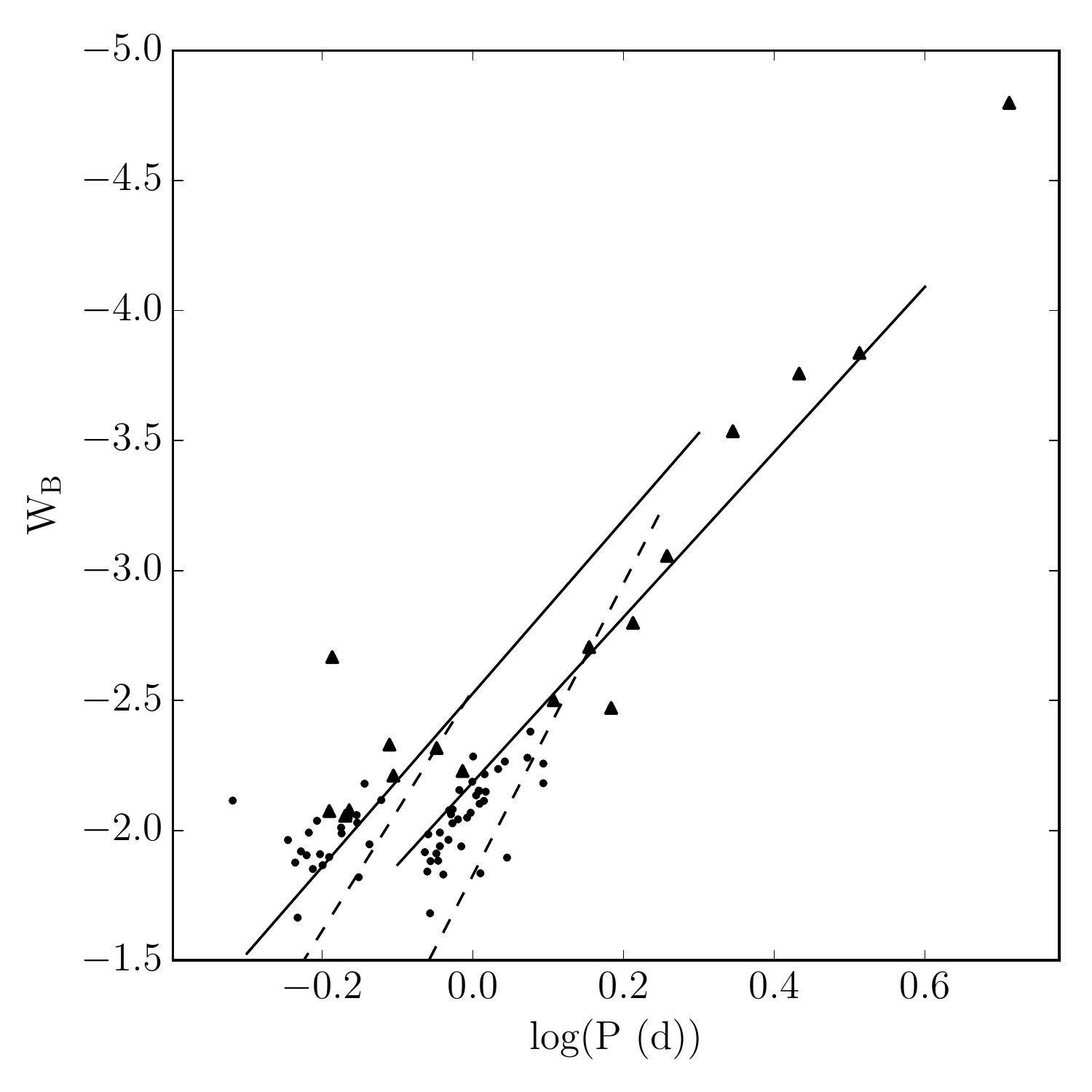}
  \end{array}$
    \caption{\small{\emph{Left:} The $B$-band PL relation for the
    Cepheids in DDO210. The solid lines are the PL relations for
    fundamental and first-overtone, short-period CCs from
    \citet{gal04}. The dashed lines are the PL relations for ACs from
    \citet{pri02}.  Symbols for the points are as in Fig.
    \ref{fig:ceph_cmd}. \emph{Right:} The Wesenheit PL relation for
    the Cepheids in DDO210.\label{fig:bwpl}}}
\end{figure*}

The locations of these stars in the CMD coupled with the AMR from
\citet{col14} indicate that most if not all of these stars are in fact
CCs. In an attempt to place furter constraints on the nature of these
Cepheids, we show the $M_B$ and reddening-free, Wesenheit PL relations
in Fig. \ref{fig:bwpl}. Again, we have used the RR Lyrae distance
modulus and extinctions as before to convert to absolute
magnitudes. We have taken several PL relations from the literature for
CCs and ACs to compare with the Cepheids in DDO210. For the CCs, we
use the $B$-band PL relation from \citet{gal04} for short-period
CCs. As for the ACs, we use the $B$-band PL relation from
\citet{pri02}. Although there is a considerable amount of scatter, it
appears that the likely CCs identified using Fig. \ref{fig:ceph_cmd}
(triangles in Fig. \ref{fig:bwpl}; see Section \ref{sub:evo})
do follow the CC PL relations. As for the fainter Cepheids, the
picture is again unclear. The faintest of these do seem to aggregate
near the AC PL relations, but there is enough scatter to confuse the
distinction for most of them. 

The Wesenheit magnitudes inherently remove the effects of reddening to
stars \citep{bro80}. To
calculate Wesenheit magnitudes, we use the following equation:
\begin{equation}\label{eq:wes}
W_B = M_B - R(B-I)_0
\end{equation}
Here, $R$ is the ratio of total-to-selective absorption, $A_B / E(B-I)
= 1.710$ \citep{sf11}. We have constructed $W_B$ PL relations using
the $B$-band PL relations previously mentioned combined with the
$I$-band relations of \citet{fio06} for the ACs, \citet{uda99} for the
fundamental mode CCs, and \citet{sos08} for
the first-overtone CCs. We note that we needed to place the
first-overtone CC PL relation on the same scale as the fundamental mode
one by subtracting 18.449 mag from the first-overtone CC PL relation of
\citet{sos08}. This was necessary because this PL relation was not
corrected for reddening, but \citet{uda99} did not provide a
first-overtone PL relation. Thus, subtracting the difference between
the fundamental mode zero-points of \citet{sos08} and \citet{uda99}
places the first-overtone $M_I$ PL relation on the same scale as the
fundamental mode one. To construct the Wesenheit PL relation for the different
types of Cepheids, we took the $B$ and $I$ PL relations at a given
period, and calculated $W_B$ using Equation (\ref{eq:wes}). 

The result, shown in the right panel of Fig. \ref{fig:bwpl}, has clearly
reduced the scatter in the DDO210 Cepheid PL relation. While these two
PL relations lie close in this Wesenheit plane, it appears that most
of the Cepheids lie brighter than the AC PL relations and closer to
the CC PL relations in this plane. Additionally, we observe the
possible AC candidates (circles) as determined by the CMD analysis in
Section \ref{sub:evo} to lie on the
same PL relation for the likely CCs (stars) extended to fainter
magnitudes. This would seem to support this Cepheid population
containing mostly, if not entirely, CCs. Finally, we draw attention to
the two bright outliers on the short-period end of the $W_B$ PL relation. These
stars are more likely second-overtone Cepheids since these pulsators
have been observed to pulsate with shorter periods at a given
magnitude when compared to the first-overtone pulsators.

\subsection{Specific frequency of Anomalous Cepheids}
\label{sub:ac_sf}
The final test we perform to discern the nature of the Cepheids in
DDO210 is to examine the specific frequency of the potential
ACs. \citet{mat95} first noted that the specific frequency of ACs
(number per 10$^5$ L$_{V, 
\sun}$) is strongly correlated with both the absolute visual magnitude
and mean metallicity of MW satellite dSph. \citet{pri04, pri05a}
further extended the study of ACs to dSph orbiting M31 and discovered
those galaxies to follow the same trends. \citet{pri05a} also note that
Phoenix, a dTrans, follows the same trends as the dSph, indicating that
these relations may hold independent of galaxy morphological type. 

\begin{figure*}
\includegraphics[width=2\columnwidth]{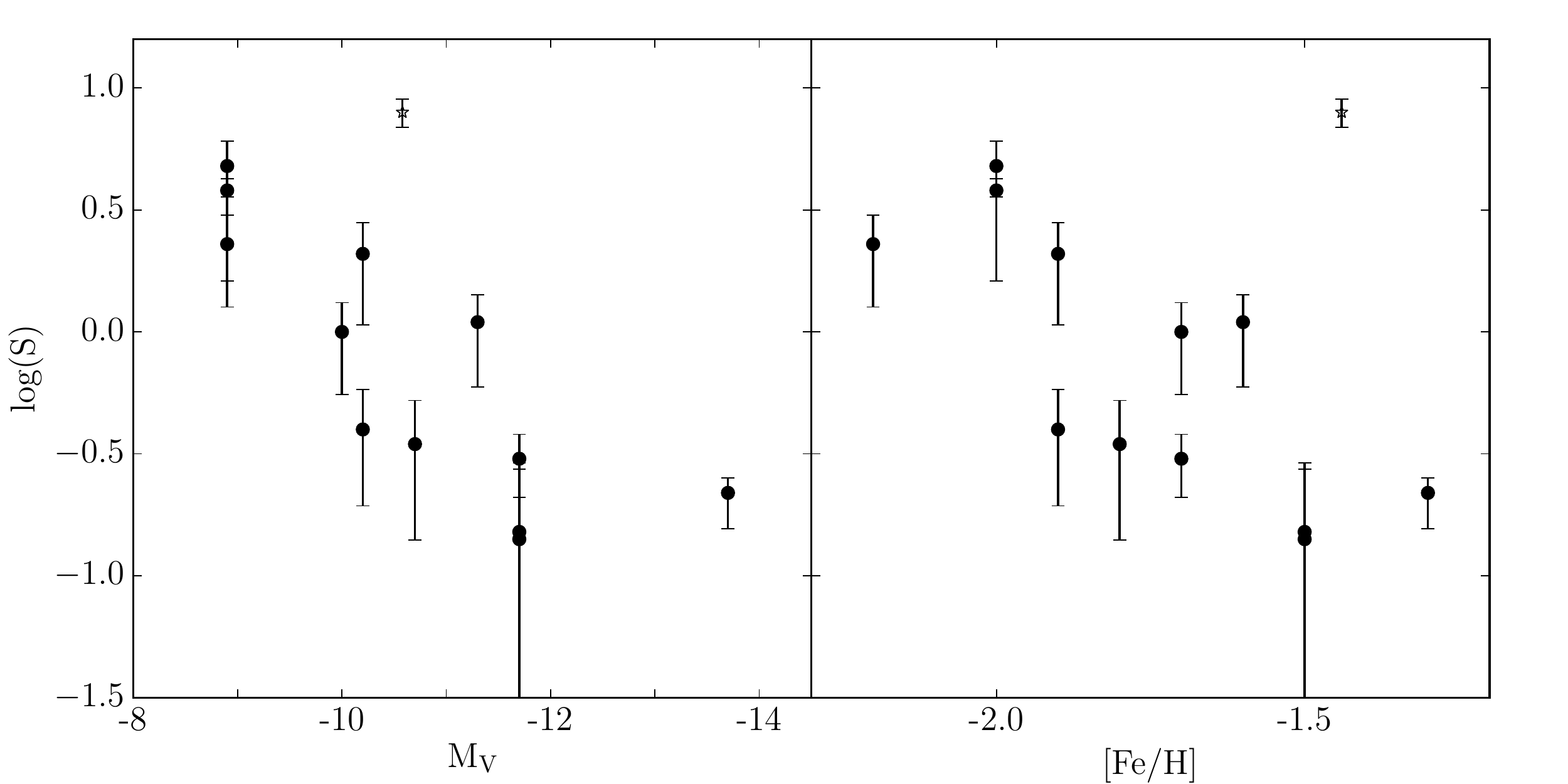}
\caption{\small{The AC specific frequency trends with absolute visual
    magnitude and mean metal abundance for LG dwarfs. Error bars are
    calculated by assuming Poisson statistics govern the errors in
    the AC counts. The open star shows where DDO210 lies if all of the
    potential ACs are true ACs. Data for other LG dwarfs from \citet{pri04, pri05a}.\label{fig:acsf}}}
\end{figure*}

If we assume that this relation holds for all dwarf galaxies in the
LG, then we expect the specific frequency of ACs in DDO210 to follow
these same trends. We have therefore calculated the AC specific
frequency in DDO210 assuming all of the 58 potential ACs identified in Section
\ref{sub:evo} are bona fide ACs. For this calculation, we have used
an absolute magnitude of $M_V = -10.58$ mag \citep{mcc06}, a surveyed
area fraction for this dataset of 50\% \citep{col14}, and mean metal
abundance of [Fe/H] $ = -1.44$ dex \citep{kir13}. We compare the
AC specific frequency of DDO210 with
the other dwarf galaxies in Fig. \ref{fig:acsf}.

Clearly, DDO210 does not follow the same AC specific frequency trends
as other LG dwarfs if we assume all potential ACs to be real. In fact,
for DDO210 to fall in line with the trend for other dwarfs, our sample
of Cepheids would need to contain $\la$2 ACs. Thus, we
are left with two scenarios: 1) Most or all of these AC candidates are
indeed true ACs, and the specific frequency trends do not hold for all
dwarfs, or 2) The specific frequency trends do hold for DDO210, and
only $\la$2 of these Cepheids are in fact ACs. Scenario 2) coupled
with the AMR of \citet{col14}, the locations of the Cepheids in the
CMD, and the Wesenheit PL relations for these Cepheids lead us to
conclude that the majority of the Cepheids in this galaxy are CCs.

\section{Discussion}
\label{sec:disc}
The main goal of this work was to investigate the properties of the
pulsating variables in the context of the SFH of DDO210. The recent
SFH analysis of \citet{col14} provides an excellent comparison to this
end. As we have already discussed, the metallicity of the RR Lyrae
stars derived in this work agrees well with the AMR from \citet{col14}
at ages of $\sim$11 Gyr, but is inconsistent with their AMR for ages
$\ga$12.5 Gyr. Interestingly, their SFH shows a deep minimum in
star formation rate at this 11 Gyr time bin. Those authors note that
this drop is robust, and that the star formation rate at this time is
only nonzero at the $\sim 2\sigma$ level. The presence of an RR
Lyrae population with a mean metallicity of $\langle$[Fe/H]$\rangle =
-1.63\pm 0.11$ confirms a weak, but nonzero star formation rate during
this early epoch in DDO210. 

We now turn to discussing our results on the context of massive
galactic halo formation. The paradigm for the formation of the MW halo
has held that a large part of it has formed through the accretion of
dwarf galaxies closely resembling those found in the LG
today. Extending this mechanism to other massive galaxies similar to
the MW, it follows that the LG dwarfs should provide excellent windows
to the systems that built many massive galaxies in general. However, as was pointed
out in Section \ref{sub:bailey}, the RR Lyrae populations in the dSph and UFDs
of the LG appear fundamentally different than those found in the MW
halo and GGCs. \citet{fio15} attribute this difference to a difference
in the metallicity of the ancient stellar populations from which the
RR Lyraes formed. They show that the dSph and UFDs did not chemically
enrich rapidly enough to produce the HASP RRab stars observed in the
Galactic halo, GGCs, and more massive dwarfs like the LMC and
Sgr. They find that galaxies need to have enriched to metallicities of
[Fe/H] $ = -1.5$ dex or more before $\sim$10 Gyr in order to produce
these HASP RRab stars.

DDO210 appears to harbor an RR Lyrae population lacking
in HASP RRab stars, similar
to the dSph and UFDs. That these stars are metal-poor of the HASP
threshold ([Fe/H] $ =
-1.63\pm0.11$ dex; see Section \ref{sub:rrab_met}) is in agreement
with the picture provided by \citet{fio15}. Those authors go on to
estimate that no more than $\sim$50\% of the Galactic halo mass could
have accreted from these low-mass, metal-poor galaxies. Therefore, it
is still possible that some galaxies resembling DDO210 could be buried
in the Galactic halo. However, it seems that the contribution of
isolated, low-luminosity dTrans like DDO210 is at a similar level to
that of the dSph and UFDs. Considering the low level of early
star-formation implied by the RR Lyrae population and the isolation of
DDO210, we conclude that galaxies resembling this one probably did not
contribute a significant amount of mass to the Galactic halo. 

\begin{figure}
\includegraphics[width=\columnwidth]{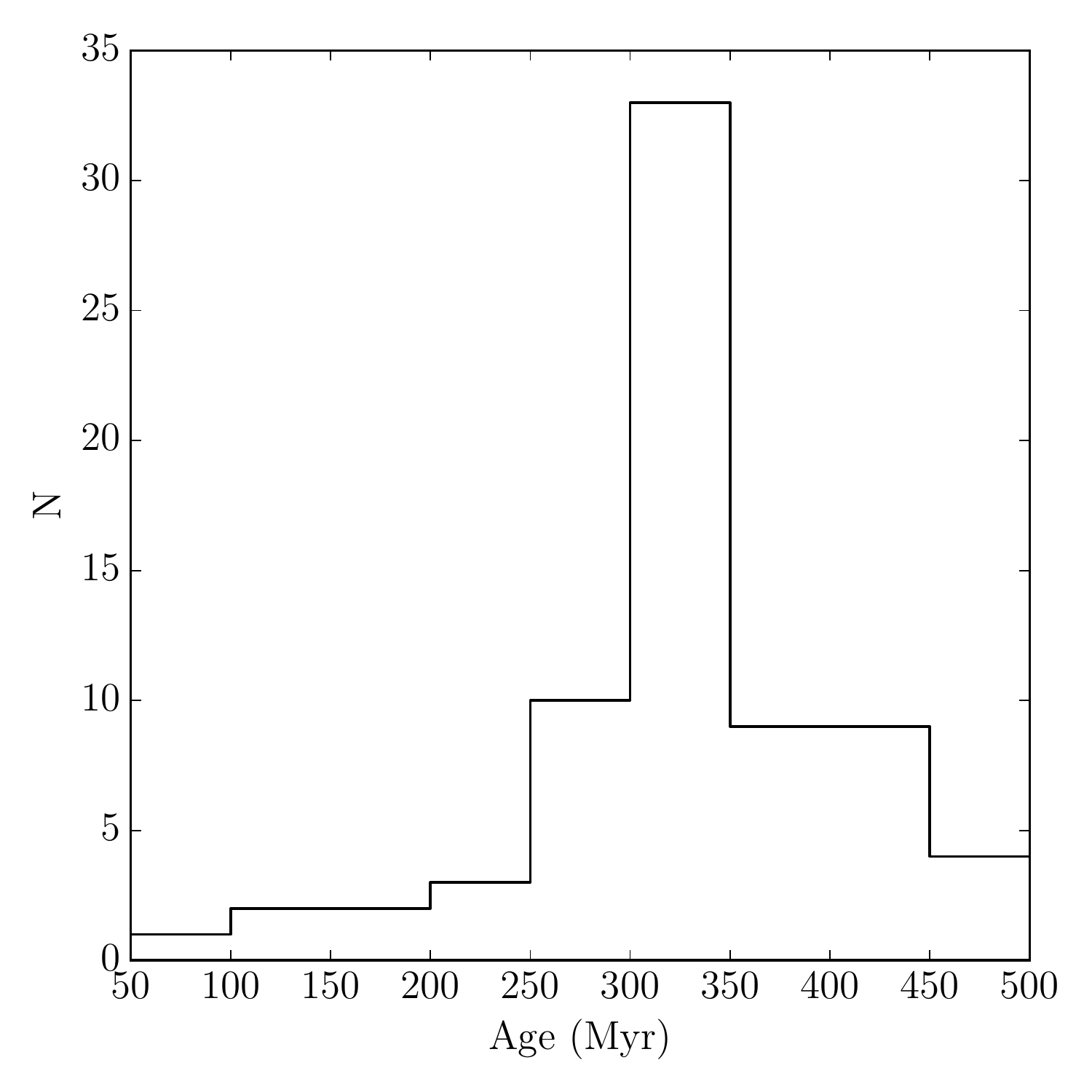}
\caption{\small{The age distribution of the Cepheid variables in
    DDO210.\label{fig:age}}}
\end{figure}

Regarding the Cepheids, it is difficult to place strong constraints on
the SFH from our sample given the unknown composition (ACs vs
CCs?). We have shown that most of this sample is likely CC, implying
that these stars formed within the past 500 Myr or so. To test this
hypothesis further, we have applied the period-age relation from
\citet{bon05} to our sample of Cepheids. Unfortunately, that study did
not extend their exploration to metallicities as low as
DDO210. Nevertheless, we continue with their most metal-poor
period-age relation ($Z = 0.004$) in an effort to gain rough insight
into the ages of these stars. The two period-age relations for the
fundamental and first-overtone Cepheids take the form:
\begin{equation}\label{eq:fupa}
\mathrm{log}\ t = 8.49 - 0.79\ \mathrm{log}\ P
\end{equation}
\begin{equation}\label{eq:fopa}
\mathrm{log}\ t = 8.41 - 1.07\ \mathrm{log}\ P
\end{equation}
Here, $t$ is the age of the Cepheid in years. We used the Wesenheit PL
relation (See Section \ref{sub:pl}) in order to distinguish between
fundamental and first-overtone Cepheids. Applying Equations
(\ref{eq:fupa}) and (\ref{eq:fopa}) to these sets of Cepheids produced
the age distribution shown in Fig. \ref{fig:age}. This distribution
indicates that, if these stars are all CCs, then these stars all
likely formed within the past Gyr. Thus, we can compare these stars'
properties with the SFH of \citet{col14} again. If we assume for the sake of
comparison that the errors introduced by using a period-age
relation for more metal-rich stars are relatively small ($\la$100 Myr), then it seems that
most of these stars were born around 300 Myr ago. \citet{col14}
examine the SFH of the past 1 Gyr in detail, and they find an
enhancement in star formation rate at ages of 250-300 Myr. Thus under
these assumptions, the properties of these CC candidates is fully
consistent with their synthetic CMD analysis. We note that these ages
are also in agreement with the analysis of \citet{mcc06} where they compare
theoretical isochrones to the position of the blue-loop stars on their
$V$, $(V-I)$ CMD. They find many of the blue-loop stars in DDO210 in the same
region as our Cepheid sample to be $\sim$300 Myr old.

\begin{figure}
\includegraphics[width=\columnwidth]{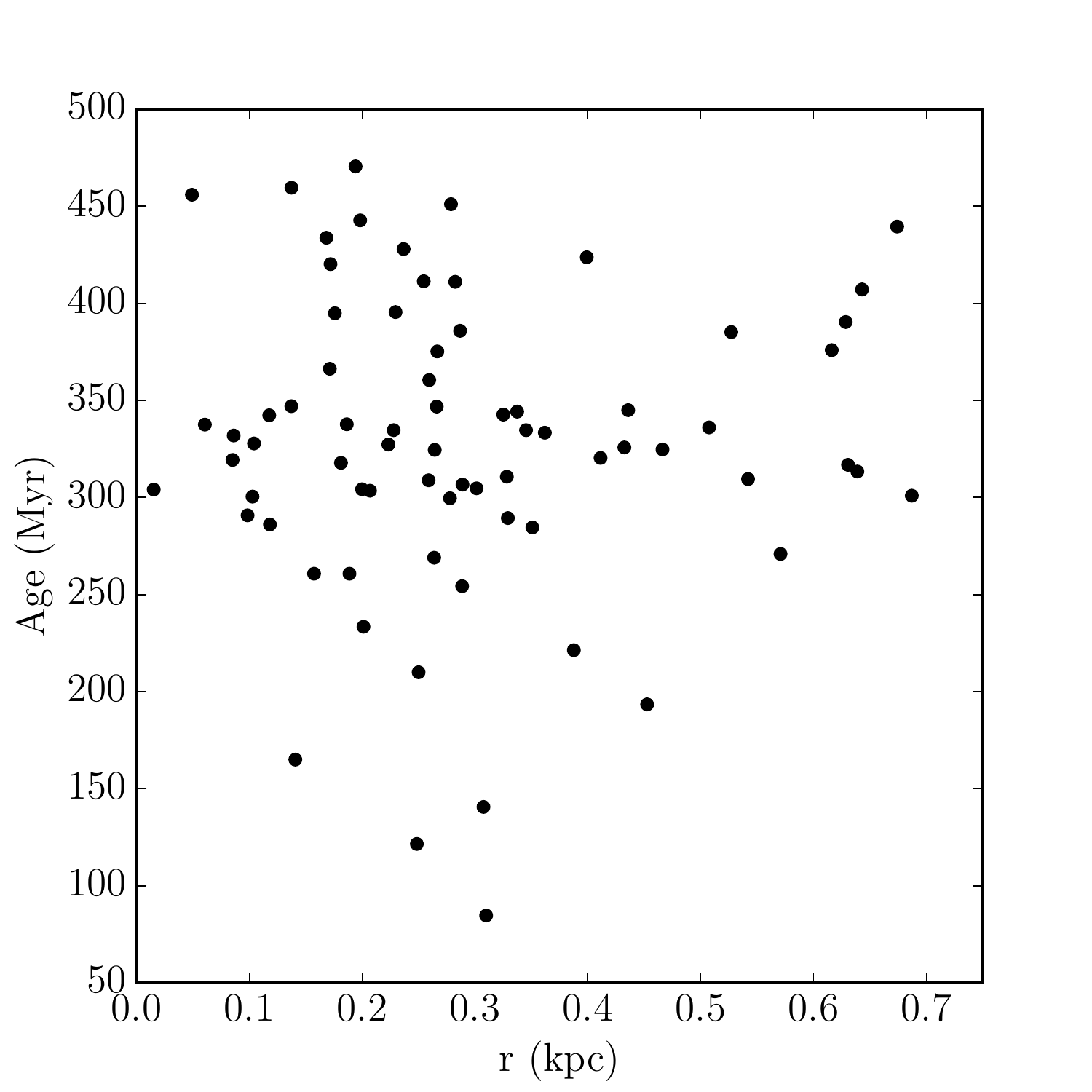}
\caption{\small{The age of the Cepheid variables in
    DDO210 versus the radial distance from the centre of the young
    stars in the galaxy. It is striking to note that no stars younger than
    $\sim$200 Myr are found beyond 0.3 kpc from the centre.\label{fig:age_grad}}}
\end{figure}

The presence of stellar population gradients has been found to be a
common characteristic of dwarf galaxies. In the case of DDO210,
\citet{mcc06} trace the distribution of different stellar populations
and find that the radial profile of the young stars is different from
the older RC and RGB stars. In particular, they find the
youngest stars to confined to a small ($\sim$0.3 kpc) clump roughly
one arcminute east of the centre of DDO210. We attempt to further examine this
young star distribution through the radial distribution of the age of the
Cepheids. In order to estimate where the clump of young stars lies in
our data, we took the centroid of all of the Cepheids younger than 250
Myr as our fiducial centre. The projected distances (assuming a
distance modulus calculated by the RR Lyrae stars) of the Cepheids
from this point are plotted against their ages in Fig. \ref{fig:age_grad}, and it reveals that the
youngest Cepheids are concentrated to within 0.4 kpc, consistent with
the results of \citet{mcc06}. On the other hand, Cepheids older than 200 Myr extend
almost twice that range in distance. 

\begin{figure}
\includegraphics[width=\columnwidth]{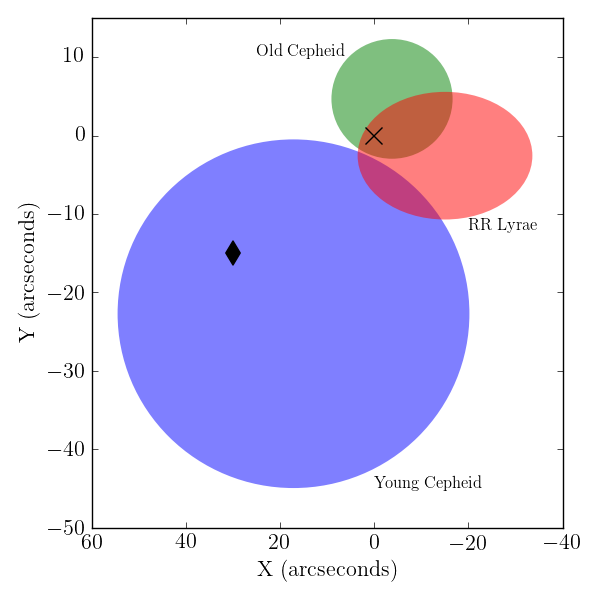}
\caption{\small{The error ellipses for the mean positions of each
    variable star population on the sky. The young Cepheids are those
    that are younger than 250 Myr, and the old Cepheids are older
    than this age. Each ellipse represents the 2 $\sigma$ error
    ellipse for each mean position in RA and Dec, where 1 $\sigma$ is
    the standard error for each mean. The black cross is the center
    of DDO210 from NED, while the black diamond is the approximate
    center of the H \texttt{I} 'dent' from \citet{mcc06}. North is
    up, east is to the left.\label{fig:var_err_ell}}}
\end{figure}

This point is further illustrated
in Fig. \ref{fig:var_err_ell}, where we have plotted the 2 $\sigma$
error ellipses for the mean positions of each variable star
population. This plot shows that even with their larger error ellipse
(owing to their small number and wide spatial distribution), the young
Cepheids are significantly offset from the center of the galaxy and
most of the other variable stellar populations. Meanwhile, the older
variables, namely Cepheids older than 250 Myr and the RR Lyrae stars,
are within 2 $\sigma$ of the center of DDO210. The young Cepheid
centroid is $\sim$ 28\arcsec ($\sim$140 pc) from the centre of
DDO210, while the older Cepheid centroid lies only
$\sim$ 6\arcsec ($\sim$30 pc) from the centre. The mean positions and
standard errors used to construct Fig. \ref{fig:var_err_ell} are
provided in Table \ref{tbl:var_xy_mean_err}.
\begin{table}
\caption{Mean positions and standard errors for each variable population.\label{tbl:var_xy_mean_err}}
\scalebox{1.}{
\begin{tabular}{ccc}
\hline
Variable type & RA (\degr) & Dec (\degr)\\
\hline
Young Cepheid & 311.7206$\pm$0.0052 & -12.8542$\pm$0.0031\\
Old Cepheid & 311.7148$\pm$ 0.0018 & -12.8466$\pm$0.0011\\
RR Lyrae & 311.7117$\pm$0.0026 & -12.8486$\pm$0.0011\\
\hline
\end{tabular}}
\end{table}

Interpretation of Fig. \ref{fig:var_err_ell} is somewhat difficult
with information available to us. This may
indicate that the recent star formation has migrated away from the
centre of the galaxy. Indeed, \citet{mcc06} already suggested that such a
scenario may have occurred within DDO210 through the process described by
\citet{doh02} for Sextans A. To summarize the mechanism, star
formation is induced on one edge of a gas cloud and thought to migrate
through the cloud from instabilities triggered by the supernova and
wind-driven shocks of the previous, adjacent star formation
episodes. The Cepheids appear consistent with such a scenario
occurring with DDO210, and in fact the centre of the youngest Cepheids
in our sample lies $\sim$ 15\arcsec ($\sim$70 pc) from the approximate
centre of the young stars and corresponding H \texttt{I} 'dent' (black diamond
in Fig. \ref{fig:var_err_ell}) discussed in
\citet{mcc06}. On the other hand, the centre of the older Cepheids in
our sample lies much further from the centre of the young stars from
\citet{mcc06} at $\sim$38\arcsec ($\sim$180 pc).

As \citet{mcc06} point out, there are other alternatives to explain
the offset of the young stars from the galactic centre. For instance,
it is possible that DDO210 interacted with some nearby system which
could have potentially induced star formation away from the
centre. However, the isolation of this galaxy renders this scenario
rather unlikely. Another
possibility that \citet{mcc06} suggest is the recent capture of
gas. However, if this mechanism was responsible for the formation of
the youngest stars, then one would not necessarily expect the
metallicity of these youngest stars to be similar to the older Cepheids
unrelated to this captured gas. The fact that these Cepheids all
appear to constitute one continuous population in the CMD and PL plane
indicate that they must be of similar chemical composition, and
therefore formed from related star formation events.

Finally, another scenario one might imagine to
explain the spatial distribution of Cepheids in this galaxy is one in
which all of these stars formed off-centre. In this case, the older
Cepheids will have had more time to migrate from the original site of
star formation. During this dispersal, the older stars have
more time to be affected by the overall galactic potential, thereby
orbiting the galactic centre as opposed to the natal star forming
region. In the absence of any kinematic information for these
stars, we cannot confirm this scenario. We do however note that recent 
theoretical work has shown that significant radial stellar migration in dwarf galaxies is
not expected to occur over these short time-scales
\citep{sch13}. 

We should also note that such
migrating star formation regions appear to occur within systems marked
by solid-body rotation \citep{doh02}. As differential rotation acts to
destroy substructure through shear, propagation of star formation
across significant distances is less likely in such systems. On the
other hand, galaxies with solid-body rotation allow for substructure
to remain coherent for longer time periods, allowing this mechanism to
more efficiently act. The galactic rotation curve for DDO210 from
\citet{beg04} shows that the central few hundred parsecs of DDO210 are likely
undergoing solid-body rotation, further supporting the possibility of
migrating star formation in this galaxy. Therefore, assuming these stars formed near where they
are presently observed, we conclude the migrating star formation
region the most likely explanation for the distribution of Cepheids in DDO210.

\section{Conclusions}
\label{sec:conc}
Using archival $HST$/ACS imaging of the LG dwarf DDO210, we have
detected over 100 pulsating variable stars within this dwarf. These
consist of 32 RR Lyrae stars and 75 Cepheids. The properties of these
pulsating variables have been compared to the SFH analysis of
\citet{col14}, showing that the SFH from \citet{col14} is consistent
with the properties of the RR Lyrae and Cepheid pulsators. In
particular, we find the relatively small population of RR Lyrae stars
to corroborate a weak but nonzero star formation rate at ages of
$\sim$11 Gyr. We find one particularly metal-poor RR Lyrae that is
also consistent with the large spread in the AMR produced by
\citet{col14} for these old ages. We find the behavior of the RR Lyrae
stars in DDO210 to be consistent with those of other LG dwarfs in the
period-amplitude plane. Specifically, this dwarf can be considered an
Oo-intermediate system with a striking lack of HASP RRab stars.

As for the Cepheids, we argue that the majority of these stars are
short-period CCs, however we cannot rule out the presence of some ACs
with the information available to us. We have utilized a period-age
relation for CCs in order to estimate the ages of these young
pulsators. We find a peak in the Cepheid age distribution near 300
Myr, which agrees well with the SFH from \citet{col14}. The youngest
of these Cepheids lie offset from the older Cepheids and the centre of
the galaxy, confirming previous studies showing the young stars lie
offset from the older stars in this galaxy. We conclude that this
offset is likely resultant from a migration of star-formation, through
a mechanism similar to what was proposed in \citet{doh02}.

\section*{Acknowledgements}

We thank Karen Kinemuchi for her comments and suggestions on an early
draft of this paper that improved its clarity. We are also grateful
for the careful reading and helpful suggestions from our referee,
Mario Mateo. This work has made use of BaSTI web tools.

\bsp	
\label{lastpage}

\begin{thebibliography}{99}
\bibitem[Alcock et al.(2000)]{alc00} Alcock, C. et al., 2000, \aj,
  119, 2194
\bibitem[Begum \& Chengalur(2004)]{beg04} Begum, A., Chengalur, J. N.,
  2004, \aap, 413, 525
\bibitem[Bernard et al.(2013)]{ber13} Bernard, E. J. et al., 2013,
  \mnras, 432, 3047
\bibitem[Brodie \& Madore(1980)]{bro80} Brodi, J. P., Madore, B. F.,
1980, \mnras, 191, 841
\bibitem[Bono et al.(2005)]{bon05} Bono, G., Marconi, M., Cassisi, S.,
  Caputo, F., Gieren, W., Pietrzynski, G., 2005, \apj, 621, 996
\bibitem[Cacciari et al.(2005)]{cac05} Cacciari, C., Corwin, T. M.,
  Carney, B. W., 2005, \aj, 129, 267
\bibitem[Catelan et al.(2004)]{cat04} Catelan, M., Pritzl, B. J.,
  Smith, H. A., 2004, \apjs, 154, 633
\bibitem[Catelan(2009)]{cat09} Catelan, M., 2009, Ap\&SS, 320, 261
\bibitem[Charbonneau(1995)]{cha95} Charbonneau, P., 1995, \apjs, 101,
  309
\bibitem[Clementini et al.(2012)]{cle12} Clementini, G., Cignoni, M.,
  Contreras Ramos, R., Federici, L., Ripepi, V., Marconi, M., Tosi,
  M., Musella, I., 2012,
  \apj, 756, 108
\bibitem[Cole et al.(2014)]{col14} Cole, A. A., Weisz, D. R., Dolphin,
  A. E., Skillman, E. D., McConnachie, A. W., Brooks, A. M., Leaman, R., 2014, \apj, 795, 54
\bibitem[Coulson \& Caldwell(1989)]{cc89} Coulson, I. M., Caldwell,
  J. A. R., 1989, \mnras, 240, 285
\bibitem[Dohm-Palmer et al.(2002)]{doh02} Dohm-Palmer, R. C.,
  Skillman, E. D., Mateo, M., et al., 2002, \aj, 123, 813
\bibitem[Dorfi \& Feuchtinger(1999)]{df99} Dorfi, E. A., Feuchtinger,
  M. U., 1999, \aap, 348, 815
\bibitem[Fiorentino et al.(2006)]{fio06} Fiorentino, G., Limongi, M.,
  Caputo, F., Marconi, M., 2006, \aap, 460, 155
\bibitem[Fiorentino \& Monelli(2012)]{fm12} Fiorentino, G., Monelli,
  M., 2012, \aap, 540, 120
\bibitem[Fiorentino et al.(2012)]{fio12} Fiorentino, G. Stetson,
  P. B., Monelli, M., Bono, G., Bernard, E. J., Pietrinferni, A., 2012,
  \apjl, 759, L12
\bibitem[Fiorentino et al.(2015)]{fio15} Fiorentino, G. et al., 2015,
  \apjl, 798, L12
\bibitem[Gallart et al.(2004)]{gal04} Gallart, C., Aparicio, A.,
  Freedman, W. L., Madore, B. F., Mart\'{i}nez-Delgado, D. Stetson,
  P. B., 2004, \aj, 127, 1486
\bibitem[Glatt et al.(2008)]{gla08} Glatt, K. et al., 2008, \aj, 135, 1106
\bibitem[Ibata, Gilmore, \& Irwin(1994)]{iba94} Ibata, R. A., Gilmore, G.,
  Irwin, M., 1994, \nat, 370, 194
\bibitem[Jacobs et al.(2009)]{jac09} Jacobs, B. A., Rizzi, L., Tully,
  R. B., Shaya, E. J., Makarov, D. I., Makarova, L., 2009, \aj,
  138, 332
\bibitem[Jurcsik \& Kov\'{a}cs(1996)]{jk96} Jurcsik, J., Kov\'{a}cs, G.,
1996, A\&A, 312, 111
\bibitem[Kirby et al.(2013)]{kir13} Kirby, E. N., Cohen, J. G.,
  Guhathakurta, P., Cheng, L., Bullock, J. S., Gallazzi, A., 2013,
  \apj, 779, 102
\bibitem[Klagyivik et al.(2013)]{kla13} Klagyivik, P., Szabados, L.,
  Szing, A., Leccia, S., Mowlavi, N., 2013, \mnras, 434, 2418
\bibitem[Layden et al.(1999)]{lay99} Layden, A. C., Ritter, L. A.,
  Welch, D. L., Webb, T. M. A., 1999, \aj, 117, 1313
\bibitem[Lee, Demarque, \& Zinn(1994)]{ldz94} Lee, Y.-W., Demarque, P,
  Zinn, R., 1994, \apj, 423, 248
\bibitem[Mancone \& Sarajedini(2008)]{man08} Mancone, C., Sarajedini,
  A., 2008, \aj, 136, 1913
\bibitem[Mateo, Fischer, \& Krzeminski(1995)]{mat95} Mateo, M., Fischer, P.,
  Krzeminski, W., 1995, \aj, 110, 2166
\bibitem[Mateo(1998)]{mat98} Mateo, M., 1998, \araa, 36, 435
\bibitem[McConnachie et al.(2005)]{mcc05} McConnachie, A. W., Irwin,
  M. J., Ferguson, A. M. N., Ibata, R. A., Lewis, G. F., Tanvir, N.,
  2005, \mnras, 356, 979 
\bibitem[McConnachie et al.(2006)]{mcc06} McConnachie, A. W., Arimoto,
  N., Irwin, M., Tolstoy, E., 2006, \mnras, 373, 715
\bibitem[Nemec et al.(2013)]{nem13} Nemec, J. M. et al., 2013, \apj,
  773, 181
\bibitem[Ordo\~{n}ez et al.(2014)]{ord14} Ordo\~{n}ez, A. J., Yang, S.-C., Sarajedini,
  A., 2014, \apj, 786, 147
\bibitem[Pietrinferni et al.(2004)]{pie04} Pietrinferni, A., Cassissi,
  S., Salaris, M., Castelli, F., 2004, \apj, 612, 168
\bibitem[Pritzl et al.(2002)]{pri02} Pritzl, B. J., Armandroff, T. E.,
  Jacoby, G. H., Da Costa, G. S., 2002, \aj, 124, 1464
\bibitem[Pritzl et al.(2004)]{pri04} Pritzl, B. J., Armandroff, T. E.,
  Jacoby, G. H., Da Costa, G. S., 2004, \aj, 127, 318
\bibitem[Pritzl et al.(2005)]{pri05a} Pritzl, B. J., Armandroff, T. E.,
  Jacoby, G. H., Da Costa, G. S., 2005, \aj, 129, 2232
\bibitem[Pritzl, Venn, Irwin(2005)]{pri05b} Pritzl, B. J., Venn, K. A., \&
  Irwin, M., 2005, \aj, 130, 2140
\bibitem[Schlafly \& Finkbeiner(2011)]{sf11} Schlafly, E. F., Finkbeiner,
  D. P., 2011, \apj, 737, 103
\bibitem[Schroyen et al.(2013)]{sch13} Schroyen, J., De Rijcke, S.,
  Koleva, M. et al., 2013, \mnras, 434, 888
\bibitem[Searle \& Zinn(1978)]{sea78} Searle, L., Zinn, R., 1978,
  \apj, 225, 357
\bibitem[Sirianni et al.(2005)]{sir05} Sirianni, M. et al., 2005,
  \pasp, 117, 836
\bibitem[Smith(1995)]{smi95} Smith, H. A., 1995, Cambridge Astrophysics
  Ser. Vol. 27, RR Lyrae Stars, Cambridge Univ. Press, Cambridge
\bibitem[Soszynski et al.(2008)]{sos08} Soszynski, I. et al., 2008,
Acta Astron., 58, 163
\bibitem[Stetson(1987)]{ste87} Steson, P. B., 1987, \pasp, 99, 191
\bibitem[Stetson(1994)]{ste94} Stetson, P. B., 1994, \pasp, 106, 250
\bibitem[Stetson(2014)]{ste14} Stetson, P. B., Fiorentino, G., Bono,
  G., Bernard, E. J., Monelli, M., Iannicola, G., Gallart, C.,
  Ferraro, I., 2014, \pasp,
  126, 616
\bibitem[Szabados et al.(2012)]{sza12} Szabados, L., Klagyivik, P.,
  2012, \aap, 537, 81
\bibitem[Tanvir(1997)]{tan97} Tanvir, N. R., 1997, in  The  Extragalactic  Distance  Scale,  ed.  M.  Livio,
M. Donahue \& N. Panagia (Cambridge: Cambridge Univ. Press), 91
\bibitem[Udalski et al.(1999)]{uda99} Udalski, A., Szymanski, M.,
  Kubiak, M., Pietrzynski, G., Soszynski, I., Wozniak, P., Zebrun, K., 1999, Acta Astron.,
  49, 201
\bibitem[Venn et al.(2004)]{ven04} Venn, K. A., Irwin, M., Shetrone,
  M. D., Tout, C. A., Hill, V., Tolstoy, E., 2004, \aj, 128, 1177
\bibitem[Yang \& Sarajedini(2012)]{yng12} Yang, S.-C., Sarajedini,
  A., 2012, \mnras, 419, 1362
\end{thebibliography}
\end{document}